\documentclass[english,prb,twocolumn,showpacs,preprintnumbers,amsmath,amssymb,floatfix,superscriptaddress]{revtex4-1}
\usepackage{graphicx}
\usepackage{multirow}

\begin{document}

\title{Quasiparticle and excitonic gaps of one-dimensional carbon chains}

\author{E.\ Mostaani}
\affiliation{Department of Physics, Lancaster
  University, Lancaster LA1 4YB, United Kingdom}
\author{B.\ Monserrat}
\affiliation{TCM Group, Cavendish Laboratory,
  University of Cambridge, 19 J.\ J.\ Thomson Avenue, Cambridge CB3 0HE,
  United Kingdom}
\affiliation{Department of Physics and Astronomy,
  Rutgers University, Piscataway, New Jersey, 08854-8019, USA}
\author{N.\ D.\ Drummond}
\affiliation{Department of Physics, Lancaster
  University, Lancaster LA1 4YB, United Kingdom}
\author{C.\ J.\ Lambert}
\affiliation{Department of Physics, Lancaster
  University, Lancaster LA1 4YB, United Kingdom}

\begin{abstract}
We report diffusion quantum Monte Carlo (DMC) calculations of the
quasiparticle and excitonic gaps of hydrogen-terminated oligoynes and
polyyne. The electronic gaps are found to be very sensitive to the atomic
structure in these systems. We have therefore optimised the geometry of
polyyne by directly minimising the DMC energy with respect to the lattice
constant and the Peierls-induced carbon--carbon bond-length alternation. We
find the bond-length alternation of polyyne to be $0.136(2)$ {\AA} and the
excitonic and quasiparticle gaps to be $3.30(7)$ and $3.4(1)$ eV,
respectively.  The DMC zone-centre longitudinal optical phonon frequency of
polyyne is $2084(5)$ cm$^{-1}$, which is consistent with Raman spectroscopic
measurements for large oligoynes.
\end{abstract}

\maketitle

\section{Introduction}
Carbon is the fourth most abundant element in the universe and is fundamental
to life as we know it. Carbon exists in a number of strikingly different
forms, including famous examples such as $sp^3$-bonded diamond and
two-dimensional $sp^2$-bonded graphene.  A less well-known form of pure carbon
is polyyne, which is a one-dimensional (1D) $sp$-bonded chain of carbon atoms
with alternating single and triple bonds. The observed presence of carbon
chains in interstellar space and circumstellar shells
\cite{Kroto_1987,Duley_2000} has inspired considerable effort to synthesise
polyyne in the laboratory, leading among other things to the discovery of
fullerenes \cite{Kroto_1985}.  Recent experiments have shown that it is
possible to produce a long linear chain of more than $200$ carbon atoms inside
a protector such as a double-walled carbon nanotube (DWCNT) \cite{Shi_2015}
and also to synthesise stable oligoynes (short polyyne molecules) with up to
44 carbon atoms \cite{Chalifoux_2010} and a variety of terminal groups
\cite{Lagow_1995,Gibtner_2002,Tsuji_2002,Matsutani_2008,Mohr_2003,Samoc_2008,Chalifoux_2011}.
Polyyne is of particular interest as the ideal interconnect in single-molecule
nanoelectronic circuitry, including spintronic devices
\cite{Crljen_2007,Standley_2008,Hong_2008,Zeng_2010}, and has potential
applications in nanomechanical devices
\cite{Baughman_2006,Liu_2013,Artyukhov_2014}. Unfortunately, the
characterisation of the optical and electronic properties of polyyne continues
to present many challenges. Our aim in this work is to address the source of
experimental and theoretical discrepancies by establishing the structural and
electronic properties of polyyne with quantitative accuracy.

The band gap of polyyne is strongly dependent on the bond-length alternation
(BLA) that arises from the so-called Peierls distortion of the linear carbon
chain \cite{Peierls_1955}. A carbon chain has a half-filled band structure
with degenerate $\pi$ orbitals; therefore a small distortion can reduce the
translational symmetry and introduce gaps into the energy bands at or near the
Fermi energy, thereby lowering the total energy.  Estimating the gap of
extended polyyne by extrapolating from the measured absorption spectra of
oligoynes has been attempted in several studies
\cite{Eisler_2005,Gibtner_2002,Mohr_2003,Samoc_2008,Zheng_2006,Chalifoux_2010,Dembinski_2000};
however, long oligoynes are needed to minimise the effects of terminal groups,
and the interpretation of the absorption spectra of oligoynes is not always
straightforward.  Most first-principles studies of the electronic structure of
polyyne to date are based on density functional theory (DFT) with different
exchange--correlation functionals
\cite{Weimer_2005,Yang_2006,Peach_2007,Zhang_2004,Imamura_2013}.  The local
density approximation (LDA) and Perdew--Burke--Ernzerhof \cite{pbe} (PBE)
functionals substantially underestimate the gap.  Hybrid exchange--correlation
functionals such as the Becke (three-parameter) Lee--Yang--Parr
\cite{Becke_1993,Lee_1998} (B3LYP) and Heyd--Scuseria--Ernzerhof
\cite{Heyd_2003,Heyd_2006} (HSE06) functionals, which include a fraction of
exact exchange, perform significantly better, but the predicted gaps still
underestimate the range of gaps indicated by experiment
\cite{Dembinski_2000,Gibtner_2002,Mohr_2003,Zhuravlev_2004,Eisler_2005,Zheng_2006,Samoc_2008,Chalifoux_2010}.
On the other hand, Hartree--Fock (HF) theory significantly overestimates gaps.
Post-HF quantum-chemistry methods such as M{\o}ller--Plesset second-order
perturbation theory (MP2) and coupled-cluster singles and doubles with
perturbative triples [CCSD(T)] offer a different and potentially far more
accurate theoretical approach \cite{Abdurahman_2002}; however the gap of
polyyne has to be obtained by extrapolating the gaps of small,
hydrogen-terminated oligoynes to infinite chain length, introducing
significant uncertainty into the results.  Previous theoretical studies have
reported the BLA of polyyne based on HF \cite{Abdurahman_2002,Yang_2006},
nonhybrid DFT \cite{Abdurahman_2002,Yang_2006}, hybrid DFT \cite{Yang_2006},
MP2 \cite{Poulsen_2001,Abdurahman_2002,Yang_2006}, and CCSD(T)
\cite{Abdurahman_2002,Zeinalipour_2008} calculations. However, there is no
consensus over either the BLA or the band gap of polyyne in the literature
\cite{Al-backri_2014,Lambert_2015}.

In this work, we use highly accurate quantum Monte Carlo (QMC) methods
\cite{Ceperley_1980,Foulkes_2001} to calculate ground-state and excited-state
total energies of isolated hydrogen-terminated oligoynes (C$_{2n}$H$_2$) and
supercells of polyyne subject to periodic boundary conditions.  The structure
of polyyne is defined by just two parameters, the lattice constant and the
BLA, enabling us to carry out a brute-force optimisation of the structure by
minimising the QMC total energy.  To the best of our knowledge this is the
first QMC study of polyyne.  We compare our data with experimental and
theoretical results in the literature.

The rest of this paper is organised in three sections: in Sec.\ $2$ we
describe the computational methodology.  Section $3$ contains our DFT and QMC
results for the BLA and electronic gaps of oligoynes and extended polyyne,
including the vibrational renormalisation. Finally, we present our conclusions
in Sec.\ $4$.

\section{Computational methodology}

\subsection{DFT calculations}

Our DFT calculations were performed using the \textsc{castep} plane-wave-basis
code \cite{castep}. We relaxed the geometries of hydrogen-terminated oligoynes
consisting of up to twelve pairs of carbon atoms using DFT-PBE and DFT-HSE06,
and we relaxed the geometry of extended polyyne using DFT-HSE06. The widths
and heights of our periodic unit cells were fixed at 20 Bohr radii and, for
oligoynes, the length was varied so that a constant amount of vacuum ($20$
Bohr radii) was maintained between images of the molecule. In our DFT
calculations for polyyne we used a grid of 30 $k$ points.  We used ultrasoft
pseudopotentials in our DFT-PBE calculations and norm-conserving
pseudopotentials in our DFT-HSE06 calculations.  The plane-wave cutoff energy
in our DFT geometry optimisations was 25 Ha.

The DFT-PBE zero-point energy and the DFT-LDA and DFT-PBE phonon dispersion
curves of polyyne were obtained using density functional perturbation theory
in a primitive cell with $100$ $k$ points in the Brillouin zone for both the
electronic calculation and the phonon calculation. The DFT-HSE06 zero-point
energy and phonon dispersion curve of polyyne were calculated using 32
primitive-cell $k$ points and the method of finite displacements in supercells
of up to 16 primitive cells.

\subsection{QMC calculations}

For our QMC calculations we used the static-nucleus variational and diffusion
quantum Monte Carlo (VMC and DMC) methods implemented in the \textsc{casino}
code \cite{Needs_2010}.  The DMC method has previously been used to study the
excitation energies of a variety of molecules and solids
\cite{Mitas_1994,Williamson_1998,Towler_2000,Williamson_2002,Drummond_2005_dia}.
The many-body trial wave function was composed of Slater determinants
multiplied by a Jastrow correlation factor \cite{Foulkes_2001}. We used
DFT-PBE orbitals, which were generated by \textsc{castep} using a plane-wave
cutoff energy of 120 Ha, and we used Dirac--Fock pseudopotentials
\cite{Trail_2005a,Trail_2005b}. The plane-wave orbitals were re-represented in
a blip (B-spline) basis before they were used in the QMC calculations
\cite{Alfe_2004}, allowing the use of aperiodic (for oligoynes) and 1D
periodic (for polyyne) boundary conditions in our QMC calculations.

\begin{table*}[!ht]
\small
\caption{Number of MD terms and orbital occupancies in each determinant for
  the neutral ground state, singlet and triplet excited states, cationic state,
  and anionic state in each of our calculations. ``H'' and ``L'' denote the
  HOMO and LUMO, respectively.  Note that the HOMO and HOMO$-1$ orbitals are
  degenerate, as are the LUMO and LUMO$+1$ orbitals.  All orbitals up to the
  HOMO$-2$ are occupied in each determinant.
\label{table:md_expansion}}
\centering
\begin{tabular}{lccccccccc}
\hline
& & \multicolumn{8}{c}{Orbital occupancy} \\
State & No.\ determinants &
\multicolumn{4}{c}{Spin-up} & \multicolumn{4}{c}{Spin-down} \\ 
& & H$-1$ & H & L & L$+1$ & H$-1$ & H & L & L$+1$ \\
\hline
Neutral ground state & 1 & $\bullet$ & $\bullet$ & & & $\bullet$ & $\bullet$ &
& \\ 
\hline
\multirow{8}{*}{Singlet excited state} & \multirow{8}{*}{8}& 
$\bullet$ & & $\bullet$ & & $\bullet$ & $\bullet$ & & \\
      &         &  & $\bullet$ & $\bullet$ & & $\bullet$ & $\bullet$ & & \\
      &         & $\bullet$ & & & $\bullet$ & $\bullet$ & $\bullet$ & & \\
      &         &  & $\bullet$ & & $\bullet$ & $\bullet$ & $\bullet$ & & \\
      &         & $\bullet$ & $\bullet$ & & & $\bullet$ & & $\bullet$ & \\
      &         & $\bullet$ & $\bullet$ & & &  & $\bullet$ & $\bullet$ & \\
      &         & $\bullet$ & $\bullet$ & & & $\bullet$ & & & $\bullet$ \\
      &         & $\bullet$ & $\bullet$ & & &  & $\bullet$ & & $\bullet$ \\
\hline
\multirow{4}{*}{Triplet excited state}   &  \multirow{4}{*}{4}&
 $\bullet$ & & & & $\bullet$ & $\bullet$ & $\bullet$ & \\
      &         &  & $\bullet$ & & & $\bullet$ & $\bullet$ & $\bullet$ & \\
      &         & $\bullet$ & & & & $\bullet$ & $\bullet$ & & $\bullet$ \\
      &         &  & $\bullet$ & & & $\bullet$ & $\bullet$ & & $\bullet$ \\
\hline
\multirow{2}{*}{Cationic state}   &  \multirow{2}{*}{2}&
 $\bullet$ & & & & $\bullet$ & $\bullet$ & &  \\
      &         &  & $\bullet$ & & & $\bullet$ & $\bullet$ & &  \\ 
\hline
\multirow{2}{*}{Anionic state}   &  \multirow{2}{*}{2}&
$\bullet$ & $\bullet$ & $\bullet$ & & $\bullet$ & $\bullet$ & &  \\
      &         & $\bullet$ & $\bullet$ & & $\bullet$ & $\bullet$ & $\bullet$ & & \\
\hline 
\end{tabular}
\end{table*}

For each oligoyne the DFT highest occupied molecular orbital (HOMO) and
HOMO$-1$ are degenerate, as are the lowest unoccupied molecular orbital (LUMO)
and LUMO$+1$. We have therefore studied the effect of multideterminant (MD)
Slater--Jastrow trial wave functions for excited, cationic, and anionic states
of oligoynes with $4$, $6$, $8$, $10$, and $24$ carbon atoms as well as a
supercell of polyyne composed of 8 primitive cells.  The Slater determinants
in the MD wave functions contained all the orbital occupancies that are
degenerate at the single-particle level. In Table \ref{table:md_expansion} we
specify the occupancy of the orbitals in the determinants used in our trial
wave functions. We used linear-least-squares energy minimisation
\cite{Nightinga_2001,Toulous_2007,Umrigar_2007} and unreweighted variance
minimisation \cite{Umrigar_1988,Drummond_2005} to optimise the MD coefficients
and the Jastrow factor, respectively. Using variance minimisation rather than
energy minimisation for the Jastrow factor improves the stability. A test for
C$_4$H$_2$ showed that the effects of additional determinants containing
promotions to the LUMO$+2$ are negligible.

The DMC energy was linearly extrapolated to zero time step and we verified
that finite-population errors in our results are negligible.  Fermionic
antisymmetry in DMC is imposed by the fixed-node approximation
\cite{Anderson_1976}, in which the nodal surface is pinned at that of the
trial wave function.  The fixed-node approximation allows us to study excited
states by using trial wave functions with the appropriate nodal topology.
Because the Jastrow factor is strictly positive, the nodal topology is purely
determined by the Slater determinants.

Twist-averaging is less important in 1D systems than two- or three-dimensional
systems; for example momentum quantisation in a 1D homogeneous electron gas
simply introduces a smooth, $O(n^{-2})$ error in the energy per particle
\cite{Lee_2011}.

\subsection{DMC quasiparticle and excitonic gaps}

A crucial quantity that characterises the electronic structure of polyyne is
the quasiparticle gap, which is the difference between the electron affinity
and the first ionisation potential.  The quasiparticle gap is the energy
required to create an unbound electron--hole pair. Within the DMC method
quasiparticle gaps are evaluated as
\begin{equation}
\Delta_{\rm qp}=E_{\rm I}-E_{\rm A}=E_++E_--2E_0,
\label{eq:gap_quasi}
\end{equation}
where $E_{\rm A}=E_0-E_+$ and $E_{\rm I}=E_--E_0$ are the electron affinity
and ionisation potential, respectively. $E_+$ and $E_-$ are the total energies
of the system with one more electron and one fewer electron, respectively,
than the neutral ground state and $E_0$ is the ground-state total energy. For
each oligoyne we separately relaxed the geometries of the neutral ground
state, the cation, and the anion using DFT-HSE06 before evaluating the DMC
ionisation potential and electron affinity and hence quasiparticle gap, i.e.,
we use the adiabatic definition of the quasiparticle gap. For polyyne, where
there are just two structural parameters, we relaxed the ground-state geometry
using DMC, and then used that geometry to obtain the vertical quasiparticle
gap; it was verified that the difference between the vertical and adiabatic
quasiparticle gaps is small for large oligoynes (see
Sec.\ \ref{sec:qp_ex_olig}). 

Excitonic gaps are evaluated as
\begin{equation}
\Delta_{\rm exc}=E_{\rm pr}-E_0,
\label{eq:gap_excitonic}
\end{equation}
where $E_{\rm pr}$ is the DMC total energy when a single electron is promoted
from the valence-band maximum to the conduction-band minimum (without changing
its spin for a singlet excitonic gap; swapping its spin for a triplet
excitonic gap). In the ground-state geometry, the singlet excitonic gap is
equivalent to the vertical optical absorption gap, i.e., the energy at which
the onset of photoabsorption occurs.  The DMC static-nucleus excitonic gaps
are corrected using the DFT vibrational renormalisation method described in
Sec.\ \ref{sub:vibration}.

The excitonic gaps are smaller than the quasiparticle gap due to the
attraction between the excited electron and the hole left in the valence band.
The exciton binding energy is the difference between the quasiparticle gap and
the excitonic gap. Fixed-node errors in the DMC total energies are always
positive and are expected to cancel to a significant extent when energy gaps
are calculated.

\subsection{Finite-size effects \label{sec:fs_effects}}

The BLA of polyyne in the ground state was evaluated for three supercells
consisting of 8, 12, and 16 primitive unit cells.  To remove finite-size
effects in the energy we fitted 
\begin{equation}
E(n)=E(\infty)+An^{-2},
\label{eq:ground_state}
\end{equation}
where $E(\infty)$ and $A$ are fitting parameters, to our DMC ground-state
energies per primitive cell $E(n)$ in supercells of $n$ primitive
cells \cite{Lee_2011}.

\begin{figure}[!ht]
\centering
\includegraphics[height=5cm]{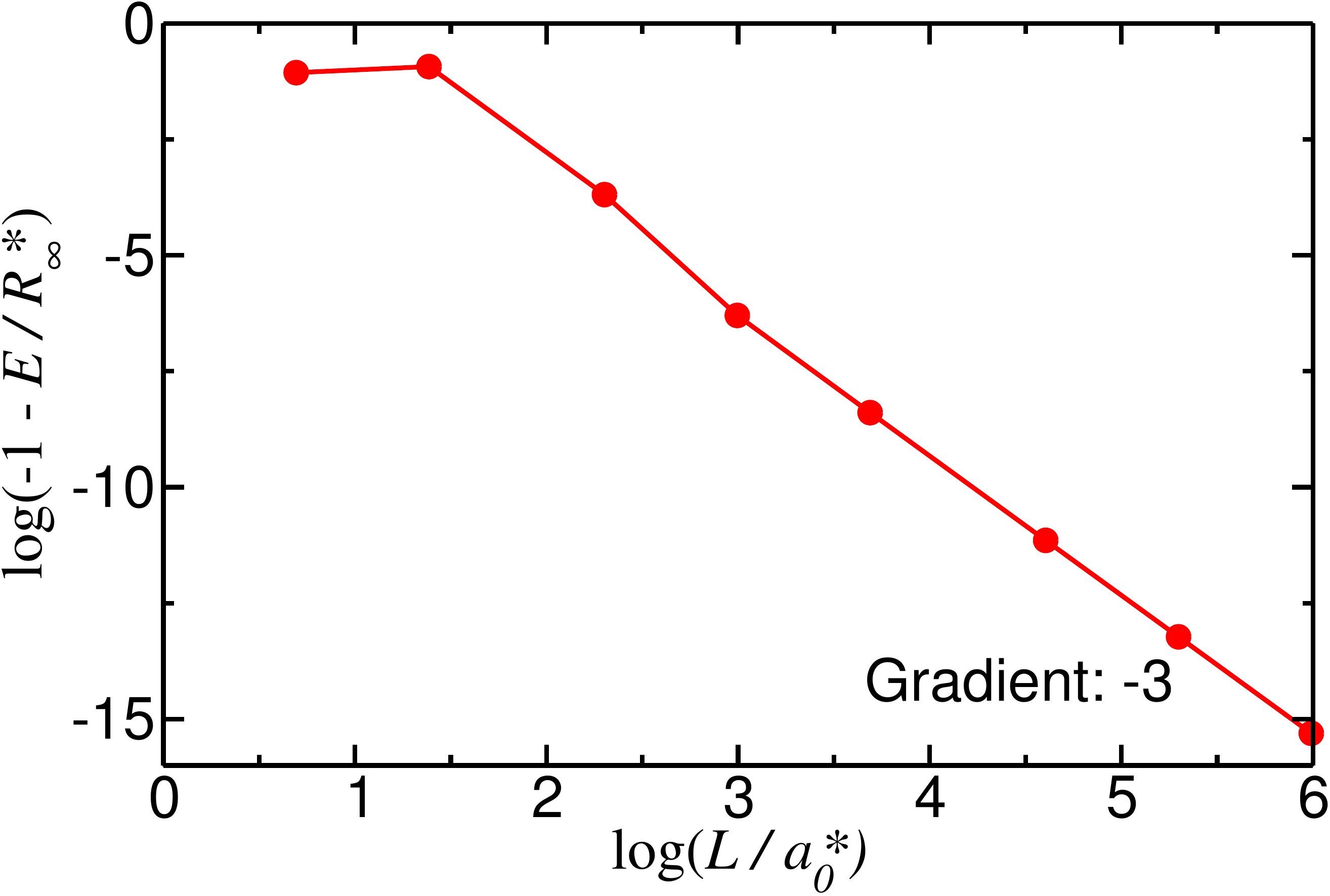}
\caption{Finite-size error in the total energy of a 1D exciton against the
  periodic cell length $L=an$, where $a$ is the lattice constant and $n$ is
  the number of primitive cells. $R^*_{\infty}=\mu/2$ is the exciton Rydberg
  and $a^*_0=1/\mu$ is the exciton Bohr radius. $\mu=m_em_h/(m_e+m_h)$ is the
  reduced mass of the electron--hole pair.
\label{fig:finite_exciton}}
\end{figure}

The DMC quasiparticle and excitonic gaps $\Delta(n)$ of polyyne were
calculated for supercells of $n=8$, $10$, $12$, and $16$ primitive cells, and
then extrapolated to infinite length by fitting
\begin{equation}
\Delta(n)=\Delta(\infty)+Bn^{-1}
\label{eq:gap_extrapolated}
\end{equation}
to the data, where $\Delta(\infty)$ and $B$ are fitting parameters.  When a
single particle is added to a finite simulation cell subject to periodic
boundary conditions, a periodic lattice of quasiparticles is formed.  The
energy of this unwanted lattice of quasiparticles goes as the Madelung
constant of the supercell lattice and results in a significant finite-size
error in the electron affinity and ionisation potential. The 1D Madelung
energy in Hartree atomic units ($\hbar=m_{\rm e}=|e|=4\pi\epsilon_0=1$) is
given by $v_M = [-0.2319-2{\rm log}(an)]/(an)$, where $a$ is the lattice
constant and $n$ is the number of primitive cells. Ignoring the logarithmic
terms, the Madelung constant falls off as the reciprocal of the linear size of
the supercell, i.e., as $1/n$. Additional finite-size effects in the exciton
energy arise from the fact that the energy is evaluated using the Ewald
interaction rather than $1/r$.  However, by calculating the ground-state
energy of an exciton modelled by a single electron and a single hole moving
strictly in 1D in a periodic cell as a function of cell length
(Fig.\ \ref{fig:finite_exciton}), we find that these finite-size errors fall
off more rapidly, as $1/n^3$. Equation (\ref{eq:gap_extrapolated}) is
therefore an appropriate fitting function for extrapolating gaps to the
thermodynamic limit.  The finite-size error in the quasiparticle gap is
significantly larger than the finite-size error in the excitonic gap, because
we do not change the number of electrons in the simulation cell when
calculating the latter.  The Madelung constant is negative, and hence the
finite-size error in the quasiparticle gap is large and negative, resulting in
a negative exciton binding energy at finite system size.  Physically this is
caused by the fact that, when a charged particle is added to or removed from a
finite, periodic cell in which particles interact via the Ewald potential, a
neutralising background is implicitly introduced.  This neutralising
background charge density vanishes in the infinite-system limit, and hence our
quasiparticle gaps are only physically meaningful in the infinite-system
limit.  For a finite molecule, by contrast, the $1/r$ Coulomb interaction is
used, and hence no additional neutralising background is introduced when a
charged particle is added to or removed from a neutral molecule.

\subsection{Vibrational renormalisation} \label{sub:vibration}

Our DMC energies have been obtained in the static-nucleus approximation. We
have used DFT methods to determine vibrational corrections to our DMC results
by including phonon zero-point energies in our reported atomisation energies
and by averaging vertical DFT excitonic gaps over phonon displacements in the
ground-state geometry to obtain a vibrational correction to the excitonic gap.

Vibrational renormalisations to electronic band gaps have recently been shown
to be as large as $-0.5$ eV for diamond
\cite{Giustino_2010,Monserrat_2014,Antonius_2014} and diamondoids
\cite{Patrick_2013}. We have therefore investigated the effects of
electron--phonon coupling on the gaps of carbon chains.

The vibrational renormalisations to the excitonic gaps were calculated at the
DFT level with the same parameters as those used for the static calculations.
Harmonic vibrational frequencies and eigenvectors were determined using the
finite-displacement method \cite{Kunc_1982}. The resulting harmonic
vibrational wave functions were used to calculate vibrational expectation
values of the gaps according to
\begin{equation}
\langle \Delta_{\rm exc}\rangle = \langle\Phi(\mathbf{q})| \Delta_{\rm
  exc}(\mathbf{q}) | \Phi(\mathbf{q})\rangle, \label{eq:zp_correction}
\end{equation}
where $|\Phi\rangle$ is the harmonic vibrational wave function and
$\mathbf{q}$ is a vector containing the amplitudes of the normal modes of
vibration, which therefore labels atomic configurations. A Monte Carlo
sampling technique \cite{Patrick_2013,Monserrat_2014_helium} was used to
evaluate Eq.\ (\ref{eq:zp_correction}). For oligoynes, a quadratic
approximation to Eq.\ (\ref{eq:zp_correction}) was also employed
\cite{Allen_1976}, yielding results consistent with those obtained
using Monte Carlo.

\subsection{Test of our method: benzene molecule}

DMC has proven to be a highly accurate method for calculating excitation
energies within the static-nucleus approximation
\cite{Mitas_1994,Williamson_1998,Towler_2000,Williamson_2002,Drummond_2005_dia}.
However, as a brief test of our methodology, we have calculated the
static-nucleus DMC ionisation potential and singlet and triplet
optical-absorption (excitonic) gaps of a benzene molecule in vacuum.  The
geometry was relaxed in both the neutral ground state and the cationic state
using DFT-PBE\@. The resulting adiabatic DMC ionisation potential is $9.24(2)$
eV, which is in excellent agreement with the experimental value of $9.24384
(6)$ eV obtained by the zero kinetic energy (ZEKE) photoelectron spectroscopy
method \cite{Nemeth_1993}. If the ground-state geometry is used for both the
ground state and the cation (i.e., the vertical ionisation potential is
calculated) then the static-nucleus DMC ionisation potential is $9.39(3)$
eV\@.  This illustrates that, when calculating ionisation potentials and
electron affinities (and hence quasiparticle gaps) for small molecules, it can
be important to relax the geometry in the neutral, cationic, and anionic
states.

Static-nucleus DMC predicts the singlet and triplet excitonic gaps of benzene
to be $5.63(4)$ and $4.56(4)$ eV, respectively, which are about 0.7 eV larger
than the experimental values of $4.9$ eV \cite{Doering_1969} and $3.9$ eV
\cite{Hiraya_1991}, respectively. This difference is largely due to the
neglect of vibrational effects.

\begin{figure}[!ht]
\centering
\includegraphics[width=.4\textwidth]{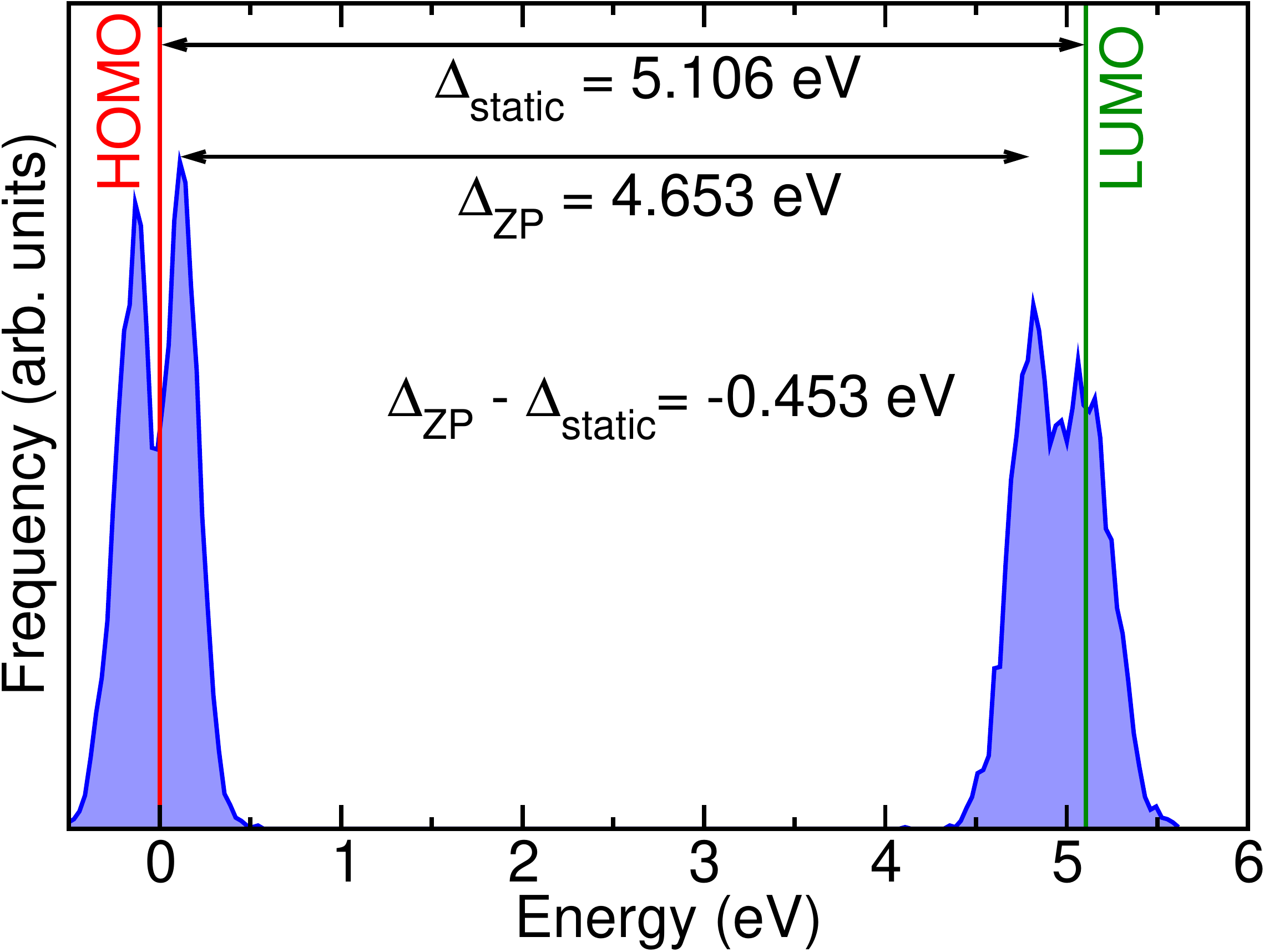}
\caption{Distribution of HOMO and LUMO DFT-PBE eigenvalues of benzene at the
  static-lattice level (vertical red and green lines) and including the
  effects of zero-point motion (shaded blue curves).
\label{fig:benzene_gap}}
\end{figure}

In Fig.\ \ref{fig:benzene_gap} we report gap-renormalisation results for
benzene, where we have relaxed the benzene molecule and calculated the band
gap using DFT-PBE\@. The static HOMO--LUMO gap is $5.106$ eV, and it reduces
to $4.653$ eV when the effects of quantum mechanical zero-point motion are
included. This gives a zero-point correction to the band gap of $-0.453$
eV\@. Using the DFT-PBE geometry, DFT-HSE06 predicts a static band gap of
$6.160$ eV, which is larger than the DFT-PBE band gap, as expected, and a
renormalised band gap of $5.660$ eV, with a zero-point correction of $-0.500$
eV\@. Similar results are obtained if the benzene molecule is relaxed using
DFT-HSE06 instead of DFT-PBE\@. We note that small changes in these results
could arise if the renormalisation were calculated for the full optical
absorption spectrum rather than individual electronic eigenvalues
\cite{Patrick_2014}.

In summary, the DFT vibrational renormalisation of the excitonic gap of
benzene ranges from $-0.45$ eV to $-0.50$ eV, depending on the choice of
exchange--correlation functional. This correction enormously improves the
agreement between theory and experiment, as previously observed in diamondoids
\cite{Giustino_2010}.  This indicates that we can expect our vibrationally
renormalised DMC gaps to be accurate to within 0.2--0.3 eV\@.

\section{Results and discussion}

\subsection{Atomic structures and atomisation energies of linear
  hydrogen-terminated oligoynes} 

\begin{figure}[!ht]
  \centering
\minipage{0.4\textwidth}
  \includegraphics[width=\textwidth]{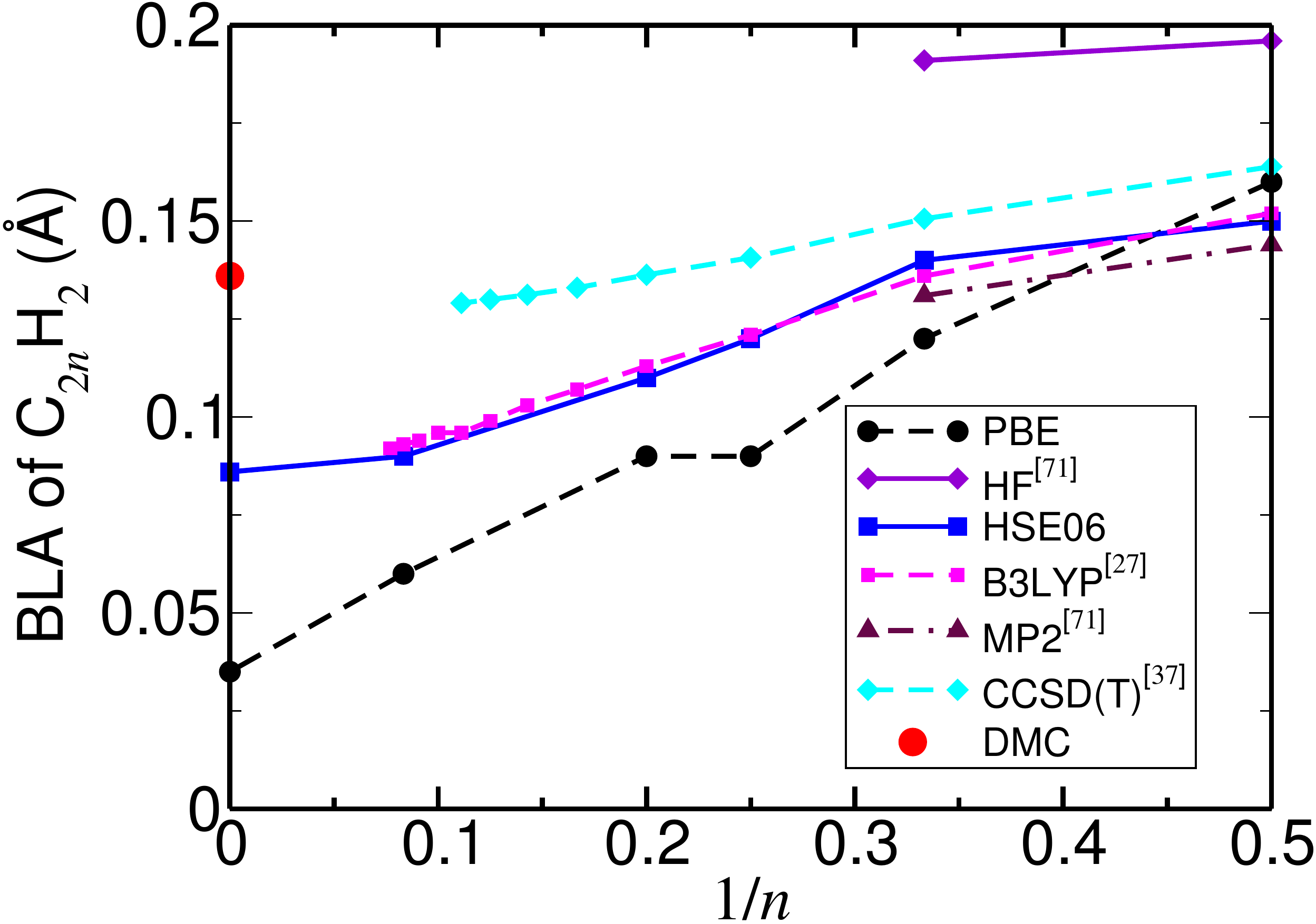}
\endminipage
\caption{Optimised BLA at the centre of a hydrogen-terminated oligoyne in
  the ground state against the reciprocal of the number $n$ of pairs of carbon
  atoms.
\label{fig:bla_oligoyne}}
\end{figure}

The ground-state BLAs at the centres of oligoynes have previously been
calculated using a variety of theoretical methods
\cite{Pino_2001,Molder_2001,Zhang_2004,Zeinalipour_2008}; some of the results
are compared with our DMC and DFT data in Fig.\ \ref{fig:bla_oligoyne}.  The
PBE functional completely fails to describe the BLA for long chains, while
spin-restricted HF theory predicts a very large BLA\@. Our DFT-HSE06 BLAs are
in agreement with the values previously obtained using the B3LYP functional
\cite{Pino_2001,Zhang_2004}, and are close to the MP2 results wherever the
latter are available \cite{Molder_2001}; however none of these BLA curves
tends to the DMC BLA of polyyne as the chain length increases.  By contrast,
the CCSD(T) BLAs \cite{Zeinalipour_2008} of oligoynes appear to tend to a
limit only slightly less than the DMC result for polyyne. Our DMC results for
the BLA of extended polyyne provide benchmark data with which the results of
other theories may be compared.
  
\begin{figure}[!ht]
\centering
 \includegraphics[width=.4\textwidth]{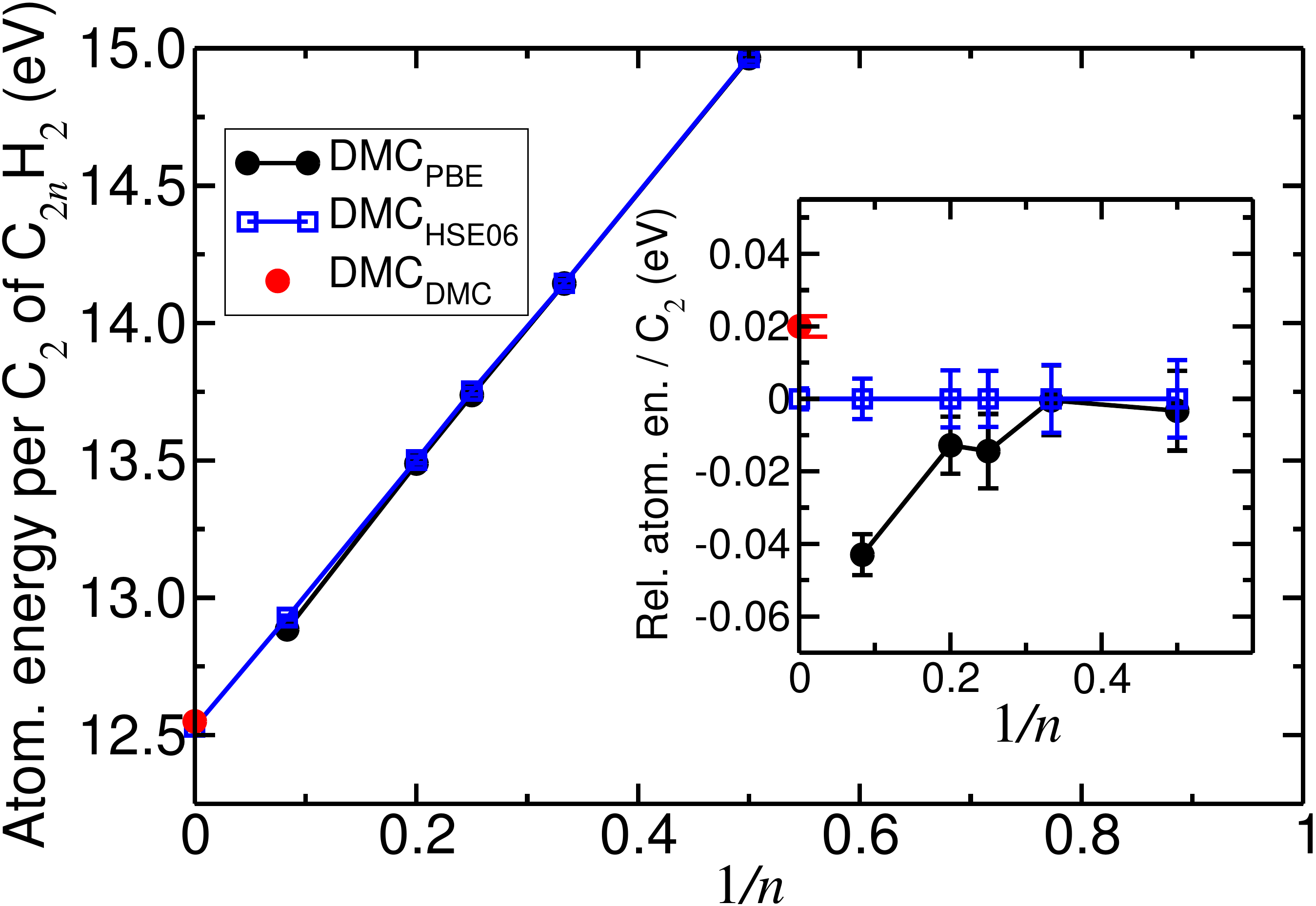}
 \caption{Static-nucleus DMC atomisation energies of hydrogen-terminated
   oligoynes as a function of the reciprocal of the number $n$ of pairs of
   carbon atoms. ``DMC$_{\rm X}$'' indicates a DMC atomisation energy
   calculated using the geometry optimised by method X\@.  The inset shows the
   relative atomisation energies of hydrogen-terminated oligoynes as a
   function of the reciprocal of the number $n$ of pairs of carbon atoms.
\label{fig:atomization_oligoyne}}
\end{figure}

The DMC static-nucleus atomisation energy of the oligoyne C$_{2n}$H$_2$ is
defined as $2n$ times the DMC total energy of an isolated, spin-polarised
carbon atom plus two times the DMC total energy of an isolated hydrogen atom
minus the DMC static-nucleus total energy of C$_{2n}$H$_2$. The DMC
atomisation energies of oligoynes obtained using geometries relaxed in
DFT-HSE06 and DFT-PBE calculations are compared in
Fig.\ \ref{fig:atomization_oligoyne}. For oligoynes consisting of up to five
pairs of carbon atoms, the difference between the DMC atomisation energies
with the DFT-PBE and DFT-HSE06 geometries is negligible.

\subsection{Atomic structure, vibrational properties and atomisation
  energy of polyyne} \label{sec:polyyne}

As the number of carbon atoms goes to infinity, the effects of the terminal
groups become negligible; therefore polyyne can be considered to be a
1D periodic chain with a primitive cell composed of two carbon
atoms with alternating triple and single bonds.

\begin{figure}[!htb]
\centering
\minipage{0.4\textwidth}
  \includegraphics[width=\textwidth]{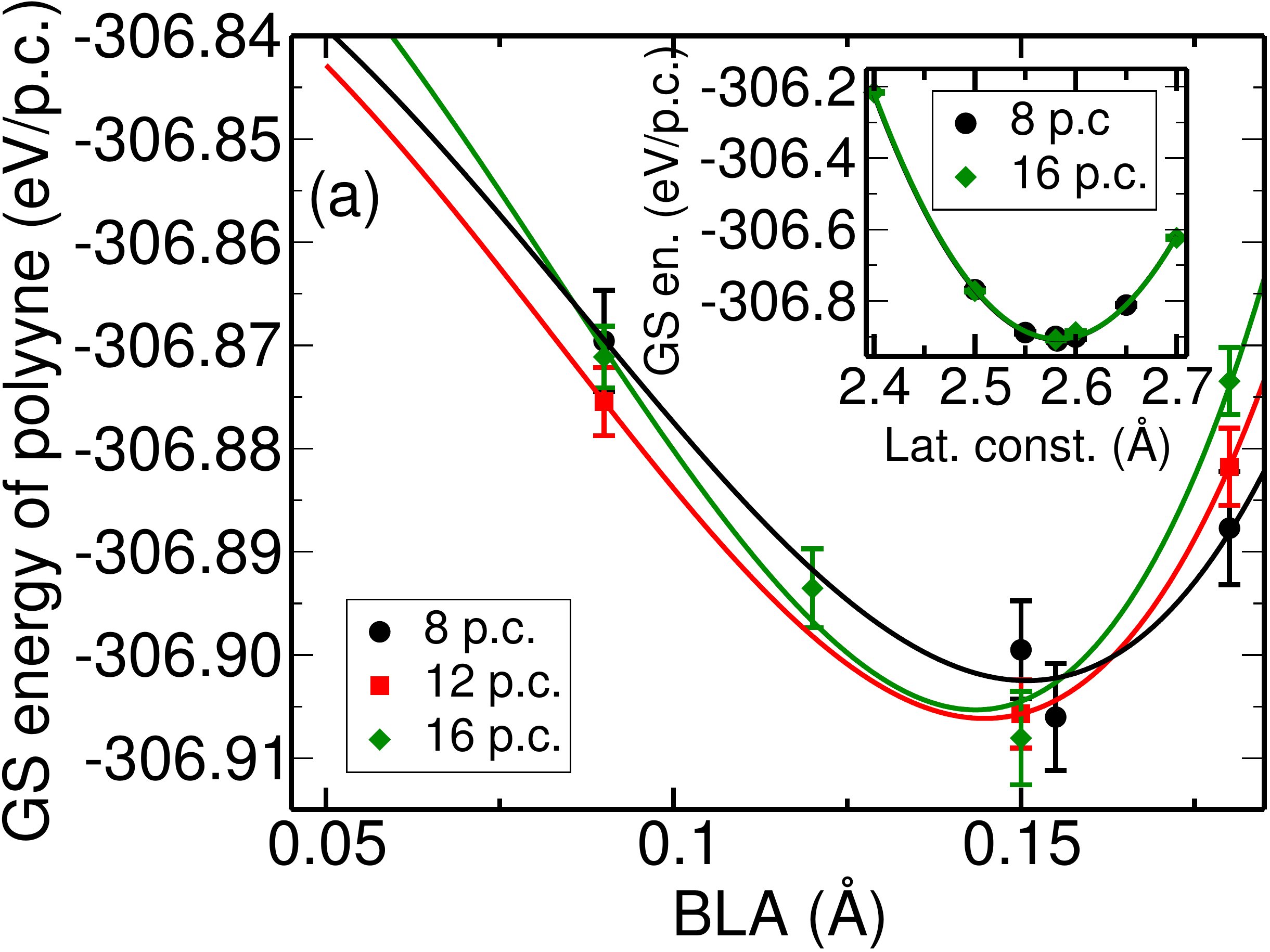}
\endminipage\\
\minipage{0.4\textwidth}
  \includegraphics[width=\textwidth]{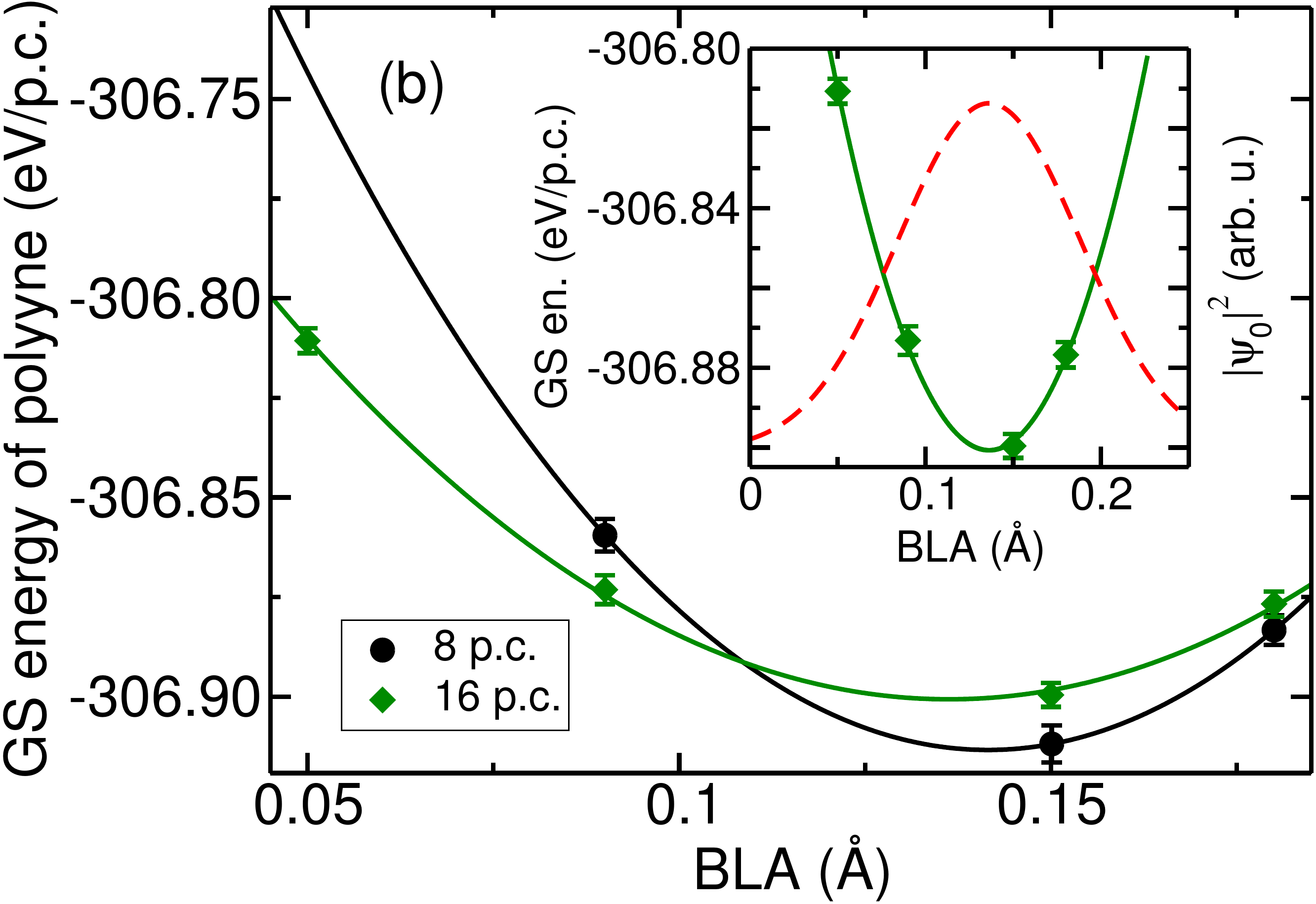}
\endminipage
\caption{(a) Ground-state (GS) DMC energy of polyyne as a function of BLA for
  lattice constant $2.58$ {\AA} in different sizes of simulation
  supercell. The inset shows the ground-state DMC energy of polyyne against
  the lattice constant at a fixed ratio of C$\equiv$C bond length to lattice
  constant for 8 primitive cells (p.c.) and a fixed C$\equiv$C bond length for
  16 primitive cells. (b) GS DMC energy of polyyne as a function of BLA for
  lattice constant $2.5817$ {\AA} in different sizes of supercell. The minimum
  of the DMC energy, $-306.901(3)$ eV per primitive cell, is at BLA $b_0=0.136
  (2)$ {\AA}. The inset shows the square modulus $|\psi_0|^2$ of the
  longitudinal optical phonon ground-state wave function for a supercell
  composed of 16 primitive cells as a function of BLA\@.
\label{fig:DMC_bla}}
\end{figure}

In order to obtain the BLA of an infinite chain, we considered supercells
subject to periodic boundary conditions, in which the lattice constant was
fixed at the DFT-BLYP \cite{Yang_2006} value of $2.58$ {\AA}. We calculated
DMC energies at different BLAs ranging from $0.09$ to $0.18$ {\AA} and fitted
a quadratic to our DMC data, as shown in Fig.\ \ref{fig:DMC_bla}(a), to locate
the minimum.

The DMC energy minima of supercells consisting of 8, 12, and 16 primitive
cells are at BLAs of $0.152 (5)$, $0.145 (2)$, and $0.144 (1)$ {\AA},
respectively. When the BLA is $0.15$ {\AA}, the C$\equiv$C triple-bond length
is $1.215$ {\AA} and the ratio of the C$\equiv$C triple-bond length to the
lattice constant is $0.471$. We then computed the ground-state DMC energy of
polyyne at several lattice constants, from $2.4$ to $2.7$ {\AA}, holding the
ratio of the C$\equiv$C bond length to the lattice constant at $0.471$ for the
supercell composed of 8 primitive cells and holding the C$\equiv$C bond length
at $1.215$ {\AA} for the supercell consisting of 16 primitive cells. The
quadratic fits to the DMC data in the inset of Fig.\ \ref{fig:DMC_bla}(a) are
in good agreement, and the ground-state energy is minimised at lattice
constants of $2.5817 (9)$ {\AA} and $2.5822(5)$ {\AA} for supercells of 8 and
16 primitive cells, respectively. Finally, the DMC energy was calculated at
lattice constant $2.5817$ {\AA} for different BLAs as shown in
Fig.\ \ref{fig:DMC_bla}(b) together with quadratic fits. The DMC energy minima
for supercells consisting of 8 and 16 primitive cells occur at BLAs of $0.142
(2)$ and $0.136 (2)$ {\AA}, respectively, which are in reasonable
agreement. Furthermore, the BLA obtained in a supercell of 16 primitive cells
does not differ significantly from the BLA $0.133(2)$ {\AA} obtained by
minimising the DMC energy extrapolated to infinite system size using
Eq.\ (\ref{eq:ground_state}).  We therefore report the BLA obtained in a
supercell of 16 primitive cells [$0.136(2)$ {\AA}] as our final result.

The DMC data shown in Fig.\ \ref{fig:DMC_bla} for the ground-state energy per
primitive cell $e(b)$ against BLA $b$ can be used to calculate the
longitudinal optical (LO) phonon frequency of polyyne at $\Gamma$.  Near the
minimum of the energy we may write
\begin{equation}
e(b)=e_0+\frac{1}{2}\frac{m_{\rm C}}{2}
\omega^2\left(\frac{b}{2}-\frac{b_0}{2}\right)^2,
\label{eq:energy_omega}
\end{equation}
where $b$ is the bond-length alternation, $b_0$ and $e_0$ are constants,
$m_{\rm C}/2$ is the reduced mass of the two carbon atoms in polyyne's
primitive unit cell, and $\omega$ is the LO phonon frequency at $\Gamma$.  In
terms of the BLA $b$, the ground-state wave function of the zone-centre LO
phonon mode of polyyne in Hartree atomic units is 
\begin{equation}
\psi_0(b)=\left(\frac{m_{\rm C}\omega}{2\pi}\right)^{1/4} \exp
\left[-\frac{m_{\rm C}\omega}{2} \left(\frac{b}{2}-\frac{b_0}{2} \right)^2
  \right].
\label{eq:wavefunction}
\end{equation}
Fitting Eq.\ (\ref{eq:energy_omega}) to the static-nucleus DMC energy of a
supercell composed of 16 primitive cells of polyyne gives $\omega=2084(5)$
cm$^{-1}$. The standard deviation of $b$ in the ground state is
$\sigma_b=\sqrt{2/(m_{\rm C}\omega)}=0.052$ {\AA}.  The square modulus of the LO
phonon ground-state wave function is plotted in the inset of Fig.\
\ref{fig:DMC_bla}(b).

\begin{figure}[!ht]
\centering
  \includegraphics[height=5cm]{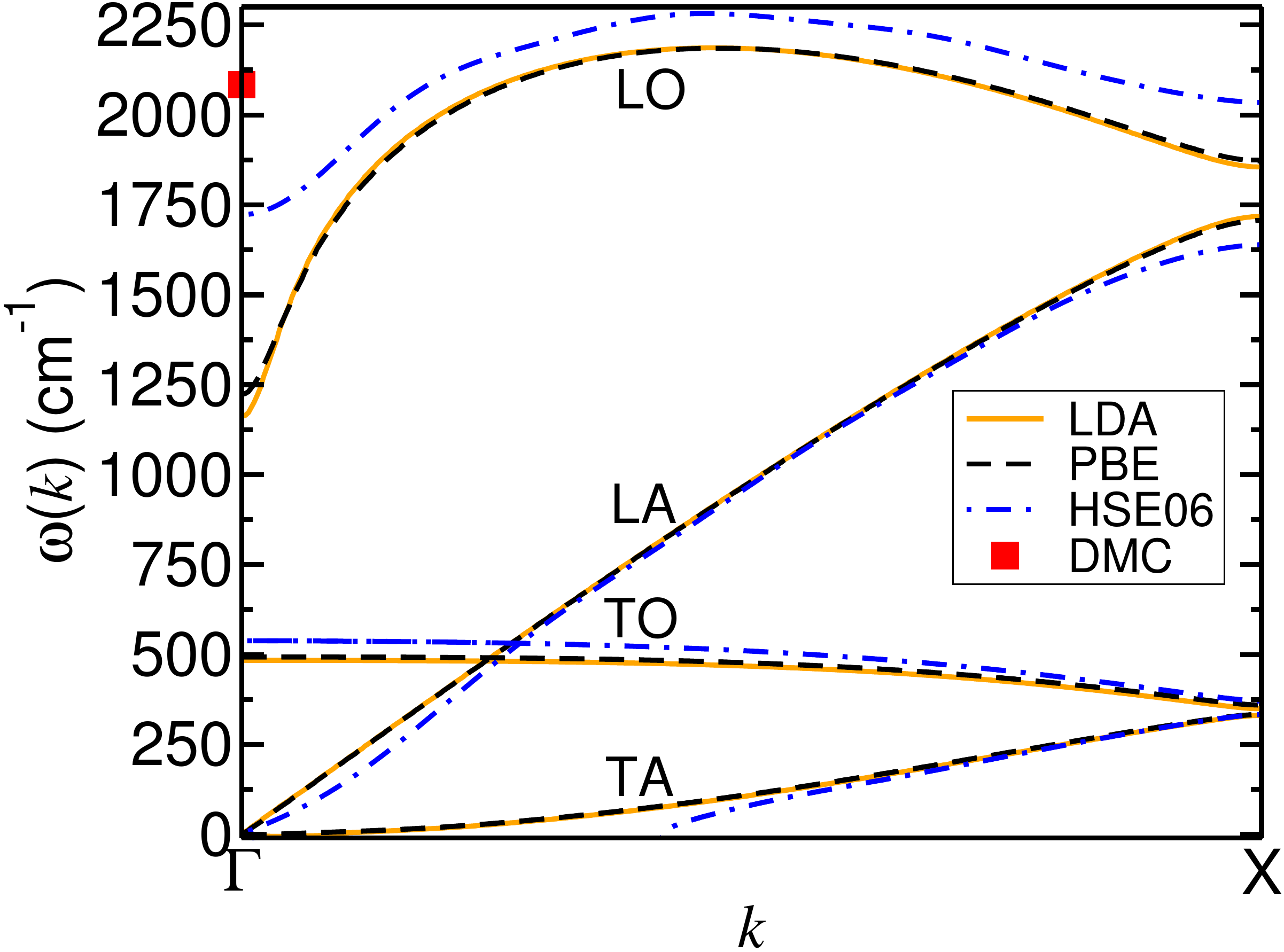}
\caption{Phonon dispersion curve of polyyne calculated using
  DFT-LDA, DFT-PBE, and DFT-HSE06. The DMC LO frequency at $\Gamma$ is shown
  by the red square. ``T,'' ``L,'' ``A,'' and ``O'' stand for transverse,
  longitudinal, acoustic, and optical, respectively. We believe the slight
  instability of the TA branch in the DFT-HSE06 dispersion curve is a
  numerical artifact.
\label{fig:chain_phonon}}
\end{figure}

In Fig.\ \ref{fig:chain_phonon} we show the DFT-LDA, DFT-PBE, and DFT-HSE06
phonon dispersion curves of polyyne. Our DFT-PBE phonon dispersion curve is in
good agreement with previous DFT-PBE results in the literature
\cite{Cahangirov_2010}. The DMC LO phonon frequency at $\Gamma$ is $2084(5)$
cm$^{-1}$, which is significantly higher than the frequencies of $1162$,
$1223$, $1723$, $1844$, and $\sim 1970$ cm$^{-1}$ obtained using DFT-LDA,
DFT-PBE, DFT-HSE06, DFT-B3LYP \cite{Milani_2008}, and equally-scaled spin
components MP2 \cite{Wanko_2016}, respectively. It is clear that DFT provides
a poor description of both the Peierls distortion and the related LO phonon
behaviour. The LO phonon frequencies of oligoynes with up to 40 carbon atoms
have been measured by Raman spectroscopy to be in the region of $1900$--$2300$
cm$^{-1}$; the precise value depends on the terminal groups, solvent, and the
number of carbon atoms in the chain \cite{Angarwal_2013}.

To evaluate the quasiparticle gap of polyyne, the atomic structure should be
in principle be relaxed when an electron is added to or removed from a
supercell.  Although the effect on the structure becomes vanishingly small as
the supercell becomes large [falling off as $O(n^{-1})$, where $n$ is the
  number of primitive cells in the supercell], the effect on the gap remains
finite, because the gap is a difference of total energies, which increase as
$O(n)$ with supercell size and depend on the atomic structure.  However, the
re-optimisation of the geometry at each system size adds noise that affects
the extrapolation to the limit of infinite system size and, as shown in
Fig.\ \ref{fig:compare_quasi}, the effect of relaxing the geometries of
cations and anions on the quasiparticle gap (i.e., the difference between the
vertical and adiabatic quasiparticle gaps) is small for large oligoynes.

In Table \ref{table:compare_bla_polyyne} we compare the equilibrium BLAs and
lattice constants of polyyne obtained using different methods. DFT-LDA, PBE,
and HSE06 functionals underestimates the BLA of polyyne, while HF theory
predicts a larger BLA than DMC\@. The DMC BLA happens to be in agreement with
the Becke--half-and-half--Lee--Yang--Parr (BHHLYP) and
Kang--Musgrave--Lee--Yang--Parr (KMLYP) results \cite{Yang_2006}. The BLA of
extended polyyne within a DWCNT has been measured to be $0.1$ {\AA}
\cite{Shi_2015}, which we expect to be different from our results for
free-standing polyyne due to the effects of charge transfer between the
polyyne and the DWCNT\@.

\begin{table}[!ht]
\small
\caption{BLA and lattice constant $a$ of polyyne as calculated or measured by
  different methods. $r_1$ and $r_2$ are the C--C and C$\equiv$C bond lengths,
  respectively. ``PBC'' indicates that periodic boundary conditions were used;
  otherwise results were obtained by extrapolation from a series of oligoynes.
  Where known, the number $n$ of pairs of carbon atoms in the longest chain
  for which calculations were performed is given. Where a citation is not
  given in the table, the data were obtained in the present work. The
  experimental result is for polyyne encapsulated in a DWCNT\@.  
\label{table:compare_bla_polyyne}}
\begin{tabular*}{0.5\textwidth}{@{\extracolsep{\fill}}lcr@{}lr@{}lr@{}lr@{}l}
\hline 
 Method  & $n$ &\multicolumn{2}{c}{$a$ ({\AA})}
&\multicolumn{2}{c}{$r_1$ ({\AA})}&
\multicolumn{2}{c}{$r_2$ ({\AA})}& \multicolumn{2}{c}{BLA ({\AA})} \\ \hline
DFT-LDA \cite{Yang_2006}  & PBC & $2.$&$566$ & $1.$&$297$ & $1.$&$269$ &
~$0.$&$028$ \\
DFT-LDA \cite{Abdurahman_2002} & PBC & $2.$&$532$ & $1.$ &$286$ & $1.$&$246$ &
$0.$&$040$ \\
DFT-PBE                  & PBC & $2.$&$565$& $1.$&$300$
&$1.$&$265$ & $0.$&$035$ \\
DFT-PBE1PBE \cite{Yang_2006} & 36 & \multicolumn{2}{c}{} & \multicolumn{2}{c}{} & \multicolumn{2}{c}{} & $0.$&$093$ \\
DFT-HSE06                 & PBC & $2.$&$56$ &$1.$&$323$ & $1.$&$237$
& $0.$&$086$ \\
DFT-KMLYP \cite{Yang_2006}  & 36 & \multicolumn{2}{c}{} & \multicolumn{2}{c}{} & \multicolumn{2}{c}{} & $0.$&$135$ \\
DFT-BHHLYP \cite{Yang_2006} & 36 & \multicolumn{2}{c}{} & \multicolumn{2}{c}{} & \multicolumn{2}{c}{} & $0.$&$134$ \\
DFT-B3LYP \cite{Yang_2006}  & 36 & \multicolumn{2}{c}{} & \multicolumn{2}{c}{} & \multicolumn{2}{c}{} & $0.$&$088$ \\
DFT-O3LYP \cite{Yang_2006}  & 36 & \multicolumn{2}{c}{} & \multicolumn{2}{c}{} & \multicolumn{2}{c}{} & $0.$&$067$ \\
DFT-BLYP \cite{Yang_2006}  & PBC & $2.$&$582$ & $1.$&$309$ & $1.$&$273$ &
$0.$&$036$ \\
HF \cite{Yang_2006}          & 36 & \multicolumn{2}{c}{} & \multicolumn{2}{c}{} & \multicolumn{2}{c}{} & $0.$&$183$ \\
MP2 \cite{Yang_2006}    & 20 & \multicolumn{2}{c}{} & \multicolumn{2}{c}{}
& \multicolumn{2}{c}{} & $0.$&$060$ \\
MP2 \cite{Abdurahman_2002} & & $2.$&$554$ &$1.$&$337$
&$1.$&$217$& $0.$&$120$ \\
MP2/CO \cite{Poulsen_2001}   & & $2.$&$6$   &$1.$&$346$ & $1.$&$254$
&$0.$&$092$ \\
CCSD \cite{Abdurahman_2002}  & &$2.$&$559$ &$1.$&$362$
&$1.$&$197$& $0.$&$165$ \\
CCSD(T) \cite{Abdurahman_2002} & &$2.$&$565$ &$1.$&$358$
&$1.$&$207$& $0.$&$151$ \\
CCSD(T) \cite{Zeinalipour_2008} & 9 &$2.$&$586$ &$1.$&$357$
&$1.$&$229$& $0.$&$128$ \\
DMC   & PBC &$2.$&$5817 (9)$ &$1.$&$359 (2)$ &$1.$&$223 (2)$ &$0.$&$136 (2)$ \\
Exp.\ in DWCNT \cite{Shi_2015}&$\sim 200$ &$2.$&$558$ &$1.$&$329$&$1.$&$229$     &$0.$&$100$ \\
\hline
\end{tabular*}
\end{table}

\begin{figure}[h]
\centering
\includegraphics[width=.4\textwidth]{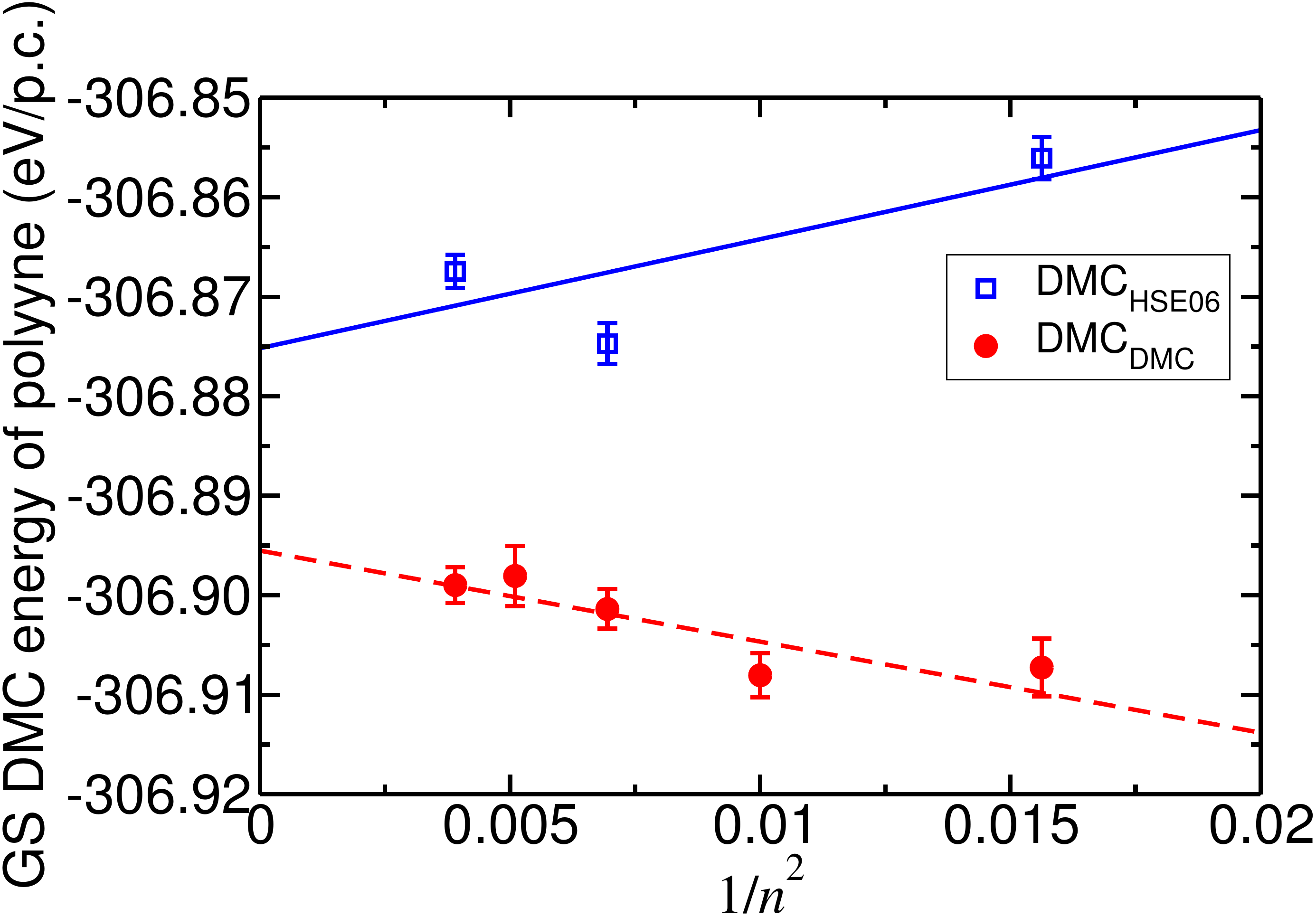}
\caption{Ground-state DMC energy of polyyne against the reciprocal of the
  square of the number $n$ of primitive cells (p.c.) in the supercell.
  ``DMC$_{\rm X}$'' indicates a DMC energy calculated using the geometry
  optimised by method X\@.
\label{fig:dmc_ground}}
\end{figure}

In Fig.\ \ref{fig:dmc_ground} we compare the ground-state DMC energy of
polyyne calculated using BLAs obtained by DMC and DFT-HSE$06$ as a function of
system size. To reduce finite-size errors, we considered supercells consisting
of 8, 12, and 16 primitive cells, with the BLA and lattice constant fixed as a
function of cell size, and we fitted a curve of the form
Eq.\ (\ref{eq:ground_state}). The extrapolated DMC energies with the DFT-HSE$06$
and DMC geometries are $-306.875 (2)$ and $-306.895 (2)$ eV per primitive
cell, respectively, confirming that DMC is needed for geometry optimisation.

DMC atomisation energies of extended polyyne obtained using DMC and DFT-HSE06
geometries are compared in Table \ref{table:compare_atomization}.  The DMC
static-nucleus atomisation energy with the DMC geometry is $12.55(1)$ eV,
which is outside the range $10.7$--$11.4$ eV estimated by MP2, CCSD, and
CCSD(T) methods in Ref.\ \citenum{Abdurahman_2002}; however the latter were
calculated by extrapolating results obtained for hydrogen-terminated oligoynes
of up to eight pairs of carbon atoms to infinite chain length, whereas our
polyyne calculations use periodic boundary conditions.  DFT phonon zero-point
energies are reported in the caption of Table
\ref{table:compare_atomization}. As shown in
Fig.\ \ref{fig:atomization_oligoyne}, the difference between DMC atomisation
energies with DFT-PBE and DFT-HSE06 geometries is negligible for small
oligoynes.

\begin{table}[!ht]
\small
\caption{Static-nucleus atomisation energy $E_c$ of polyyne as obtained by
  different methods. ``DMC$_{\rm DMC}$'' and ``DMC$_{{\rm HSE06}}$'' indicate
  that the DMC energy of polyyne was calculated using the DMC- and
  DFT-HSE06-optimised geometries, respectively. (The DFT-PBE and DFT-HSE06
  phonon zero-point energies of polyyne are $0.260$ and $0.264$ eV,
  respectively. The zero-point energy is a correction that should be
  subtracted from the atomisation energy before comparison with experiment.)
\label{table:compare_atomization}}
\centering
\begin{tabular*}{0.2\textwidth}{lr@{}l}
\hline  Method & \multicolumn{2}{c}{$E_c$ (eV)} \\ \hline
DFT-PBE                              & $13.$&$71$  \\
DFT-HSE06                            & $12.$&$47$  \\
MP2 \cite{Abdurahman_2002}            & ~~$11.$&$375$ \\
CCSD \cite{Abdurahman_2002}           & $10.$&$678$ \\
CCSD(T) \cite{Abdurahman_2002}        & $11.$&$053$ \\
DMC$_{{\rm HSE}06}$                     & $12.$&$53 (1)$ \\
DMC$_{\rm DMC}$                        & $12.$&$55 (1)$ \\
\hline 
\end{tabular*}
\end{table}

\subsection{Quasiparticle and excitonic gaps of hydrogen-terminated
  oligoynes} \label{sec:qp_ex_olig}

The DFT-HSE06 band structure of polyyne is shown in Fig.\ \ref{fig:bs_hse}.
Polyyne is a semiconductor with a direct band gap at the $X$ point of the
Brillouin zone, as expected on the basis of the Peierls distortion mechanism.

\begin{figure}[!ht]
\centering
\includegraphics[width=0.4\textwidth]{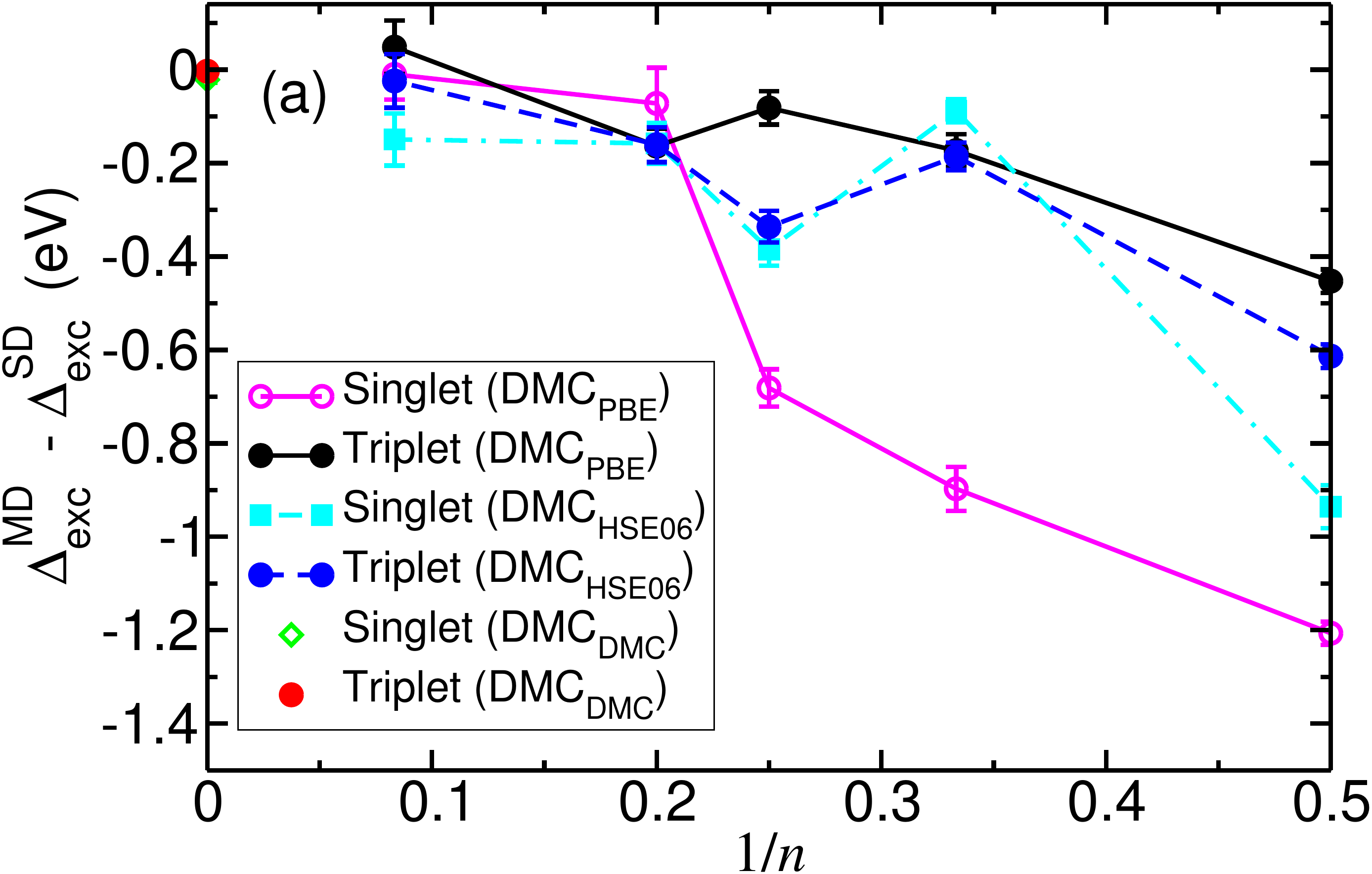} \\[2em]
\includegraphics[width=0.4\textwidth]{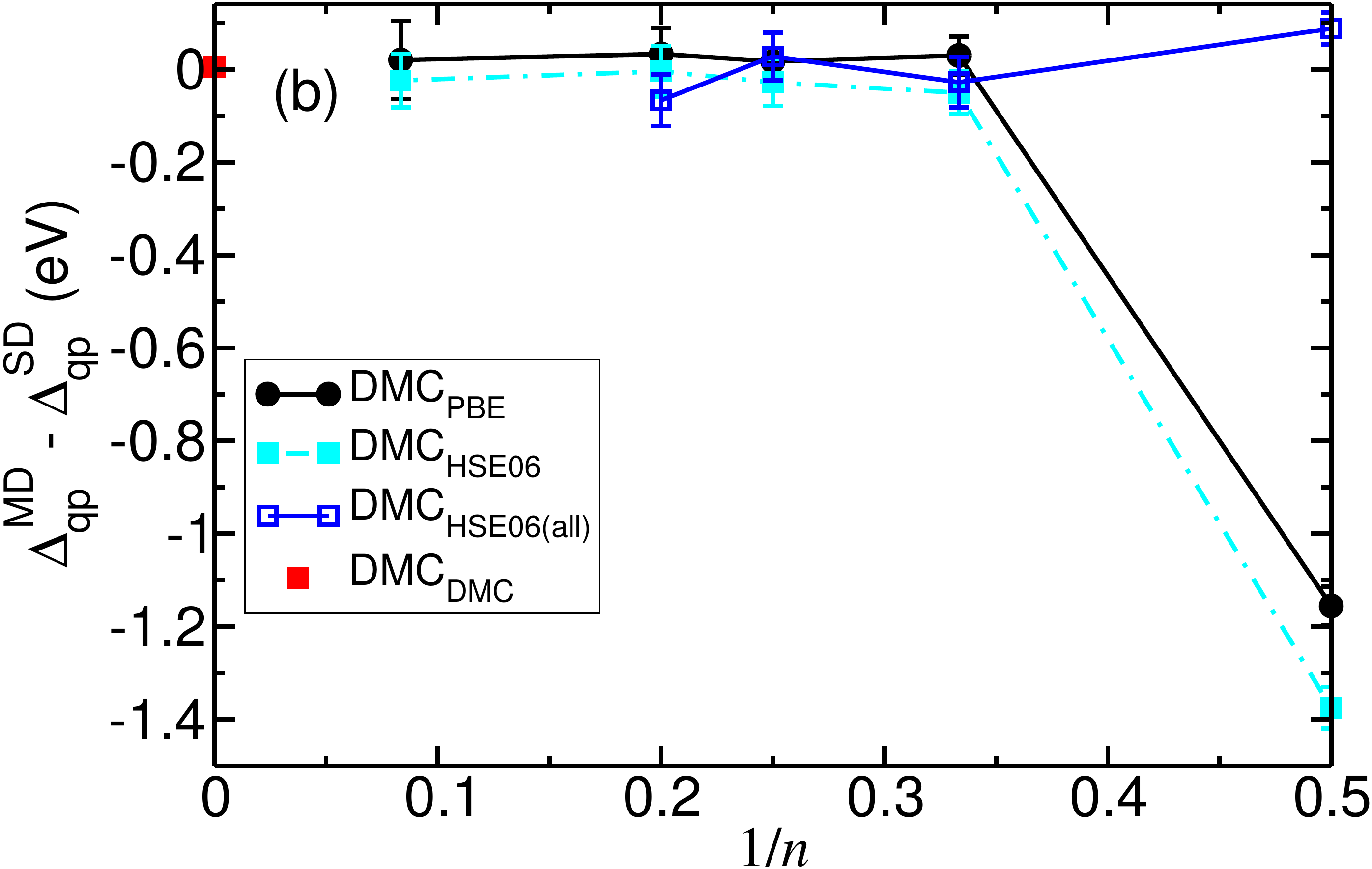}
\caption{(a) Difference ($\Delta_{\rm exc}^{\rm MD}-\Delta_{\rm exc}^{\rm
    SD}$) of the DMC excitonic gaps of oligoynes obtained using MD and
  single-determinant Slater--Jastrow trial wave functions as a function of the
  reciprocal of the number $n$ of pairs of carbon atoms. (b) Difference
  ($\Delta_{\rm qp}^{\rm MD}-\Delta_{\rm qp}^{\rm SD}$) of the DMC
  quasiparticle gaps of oligoynes obtained using MD and single-determinant
  Slater--Jastrow trial wave functions as a function of the reciprocal of the
  number $n$ of pairs of carbon atoms. DMC$_{\rm X}$ indicates a DMC gap
  calculated using the geometry optimised by method X\@. ``X(all)'' in the
  subscript indicates the use of geometries separately optimised using method
  X for the neutral ground state, cationic state, and anionic
  state.
\label{fig:md_sj}}
\end{figure}

Figure \ref{fig:md_sj}(a) shows that using an MD trial wave function reduces
the DMC singlet and triplet excitonic gaps of small oligoynes (by up to $1.3$
eV for C$_4$H$_2$).  The reduction in singlet gaps is larger than the
reduction in triplet gaps.  However, Fig.\ \ref{fig:md_sj}(b) shows that using
an MD wave function does not significantly affect the quasiparticle gaps of
oligoynes apart from C$_4$H$_2$.  As the length of the molecule increases, the
effects of using multiple determinants on the excitonic gaps decreases,
becoming negligible for polyyne.

\begin{figure}[!ht]
\centering
\includegraphics[width=.4\textwidth]{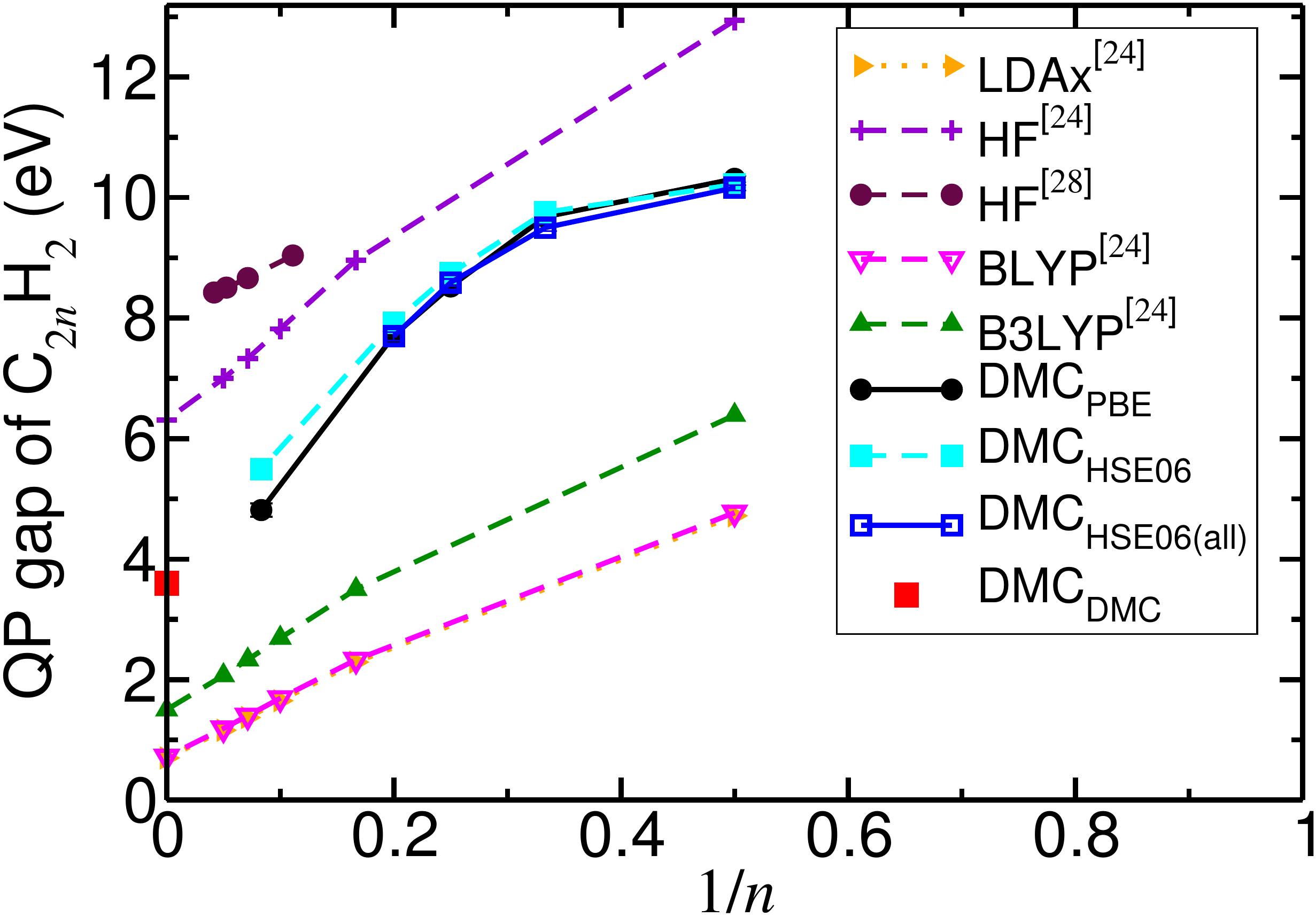}
\caption{Static-nucleus quasiparticle (QP) gaps of hydrogen-terminated
  oligoynes against the reciprocal of the number $n$ of pairs of carbon
  atoms. ``DMC$_{\rm PBE}$'' and ``DMC$_{\rm HSE06}$'' denote DMC gaps
  calculated using DFT-PBE and DFT-HSE06 ground-state geometries,
  respectively. ``DMC$_{\rm X (all)}$'' denotes DMC quasiparticle gaps
  calculated using geometries optimised by method X separately for the neutral
  ground state, cationic state, and anionic state.
\label{fig:compare_quasi}}
\end{figure}

The DMC quasiparticle gaps of oligoynes are compared with other theoretical
results in Fig.\ \ref{fig:compare_quasi}. The HF method overestimates the
quasiparticle gap, while DFT with various functionals considerably
underestimates the gap. The DMC quasiparticle gaps calculated using DFT-HSE06
and DFT-PBE geometries are in agreement for oligoynes consisting of fewer than
ten carbon atoms, but gradually start to differ from each other for longer
oligoynes, with the difference in the DMC gaps reaching $0.8(1)$ eV for
C$_{24}$H$_2$. This demonstrates that, not only the method used to calculate
the gap, but also the method used to optimise the geometry of polyyne must be
highly accurate. Using the ground-state geometry rather than separately
optimised geometries for the ground, cationic, and anionic states increases
the quasiparticle gap by less than $0.15$ eV for oligoynes longer than
C$_8$H$_2$ (i.e., the difference between the vertical and the adiabatic
quasiparticle gap is negligible for large oligoynes).  The DMC quasiparticle
gap of polyyne, evaluated using the DMC ground-state geometry, is $3.6(1)$
eV\@.

\begin{figure}[htb]
\centering
  \includegraphics[width=0.4\textwidth]{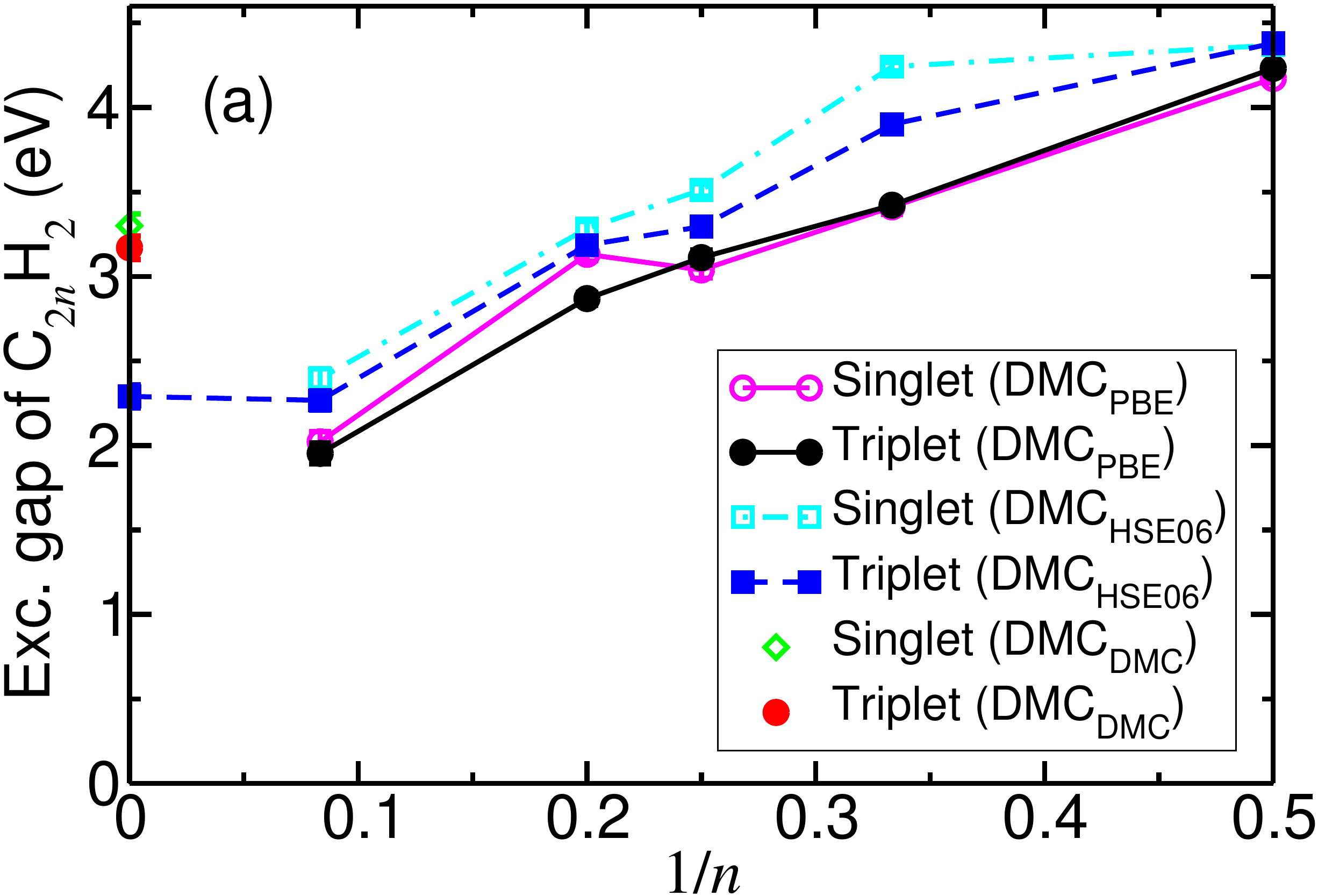}\\[2em]
  \includegraphics[width=0.4\textwidth]{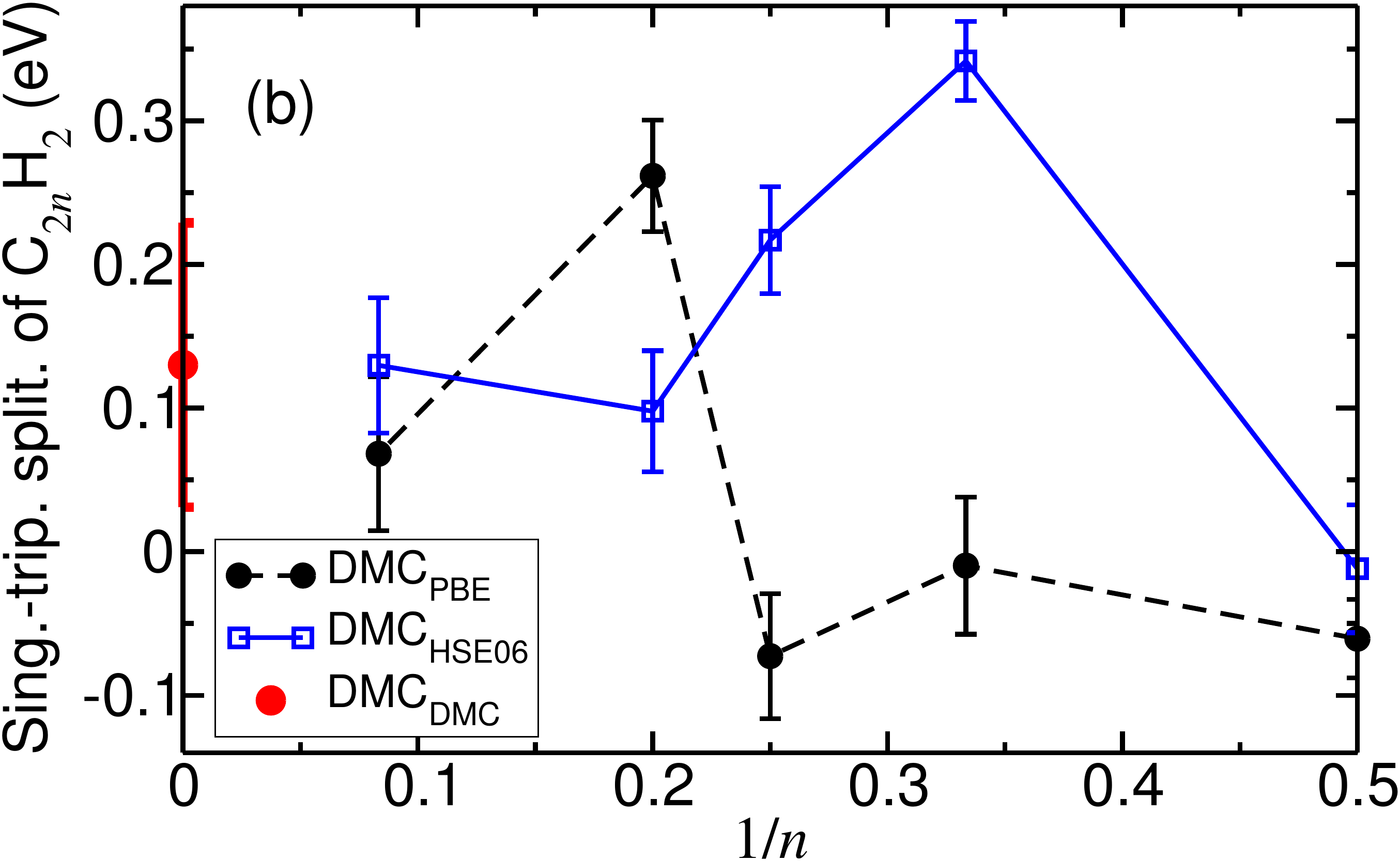}
\caption{(a) DMC static-nucleus singlet and triplet excitonic gaps for
  oligoynes, whose geometries are optimised by DFT-PBE and DFT-HSE06, against
  the reciprocal of the number $n$ of pairs of carbon atoms.  DMC$_{\rm X}$
  indicates a DMC gap calculated using the geometry optimised by method
  X\@. (b) DMC singlet--triplet splitting for oligoynes obtained with DFT-PBE
  and DFT-HSE06 geometries. The polyyne limit was obtained using the DMC
  geometry.
\label{fig:compare_gap}}
\end{figure}

We plot the static-nucleus singlet and triplet excitonic gaps of different
oligoynes in Fig.\ \ref{fig:compare_gap}(a). Singlet--triplet splitting (the
difference of singlet and triplet excitonic gaps) against the reciprocal of
the number $n$ of pairs of carbon atoms in oligoynes is small, about
$0.1$--$0.2$ eV as shown in Fig.\ \ref{fig:compare_gap}(b). Using DFT-HSE06
geometries instead of DFT-PBE geometries typically increases the DMC gaps by
around 0.2 eV for small oligoynes. The DMC singlet and triplet excitonic gaps
of extended polyyne using the ground-state DMC geometry are obtained by
extrapolating results obtained in finite, periodic cells to infinite system
size, as discussed in Sec.\ \ref{sec:gap_polyyne}.

\begin{figure}[!ht]
\centering
\includegraphics[width=.4\textwidth]{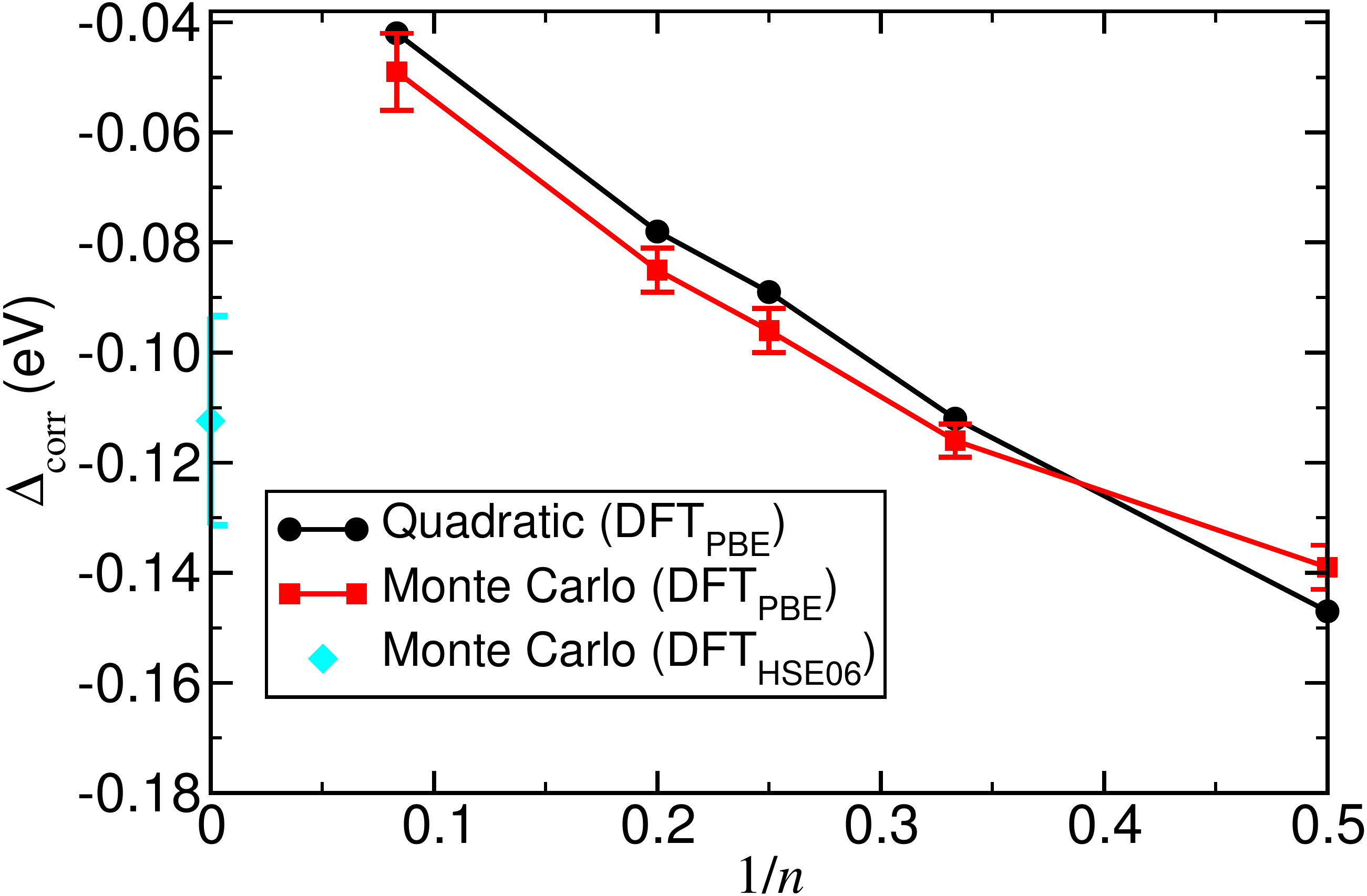}
\caption{DFT-PBE zero-point correction $\Delta_{\rm corr}$ to the excitonic
  gaps of oligoynes as a function of the reciprocal of the number $n$ of pairs
  of carbon atoms. The zero-point correction in the polyyne limit was
  calculated by DFT-HSE06.
\label{fig:oligoynes_gap}}
\end{figure}

In Fig.\ \ref{fig:oligoynes_gap} we show zero-point corrections to the
excitonic gaps of hydrogen-terminated oligoynes. The error bars in the Monte
Carlo results indicate the statistical uncertainty arising from the Monte
Carlo integration. The band-gap corrections calculated using the quadratic
method are in good agreement with the Monte Carlo results. In the quadratic
method, the coupling of each vibrational normal mode to the electronic band
extrema is treated individually, hence providing access to the microscopic
behaviour of the system. Within DFT-PBE, the largest phonon zero-point
correction to the gap is found in the shortest oligoyne considered,
C$_4$H$_2$, at about $-0.14$ eV\@. The correction decreases with increasing
chain length to about $-0.05$ eV for C$_{24}$H$_2$. The decrease in the
strength of electron--phonon coupling with increasing chain size in oligoynes
can be attributed to the decrease in the importance of the hydrogen atoms at
the terminations. The DFT-HSE06 zero-point correction to the excitonic gap of
polyyne is obtained by extrapolation to infinite system size as explained in
Sec.\ \ref{sec:gap_polyyne}.  Phonon renormalisation of gaps is clearly not as
important in oligoynes as in either benzene or diamond.

\subsection{Quasiparticle and excitonic gaps of polyyne} \label{sec:gap_polyyne}
 
\begin{figure}[!htb]
\centering
\minipage{0.4\textwidth}
\includegraphics[width=\textwidth]{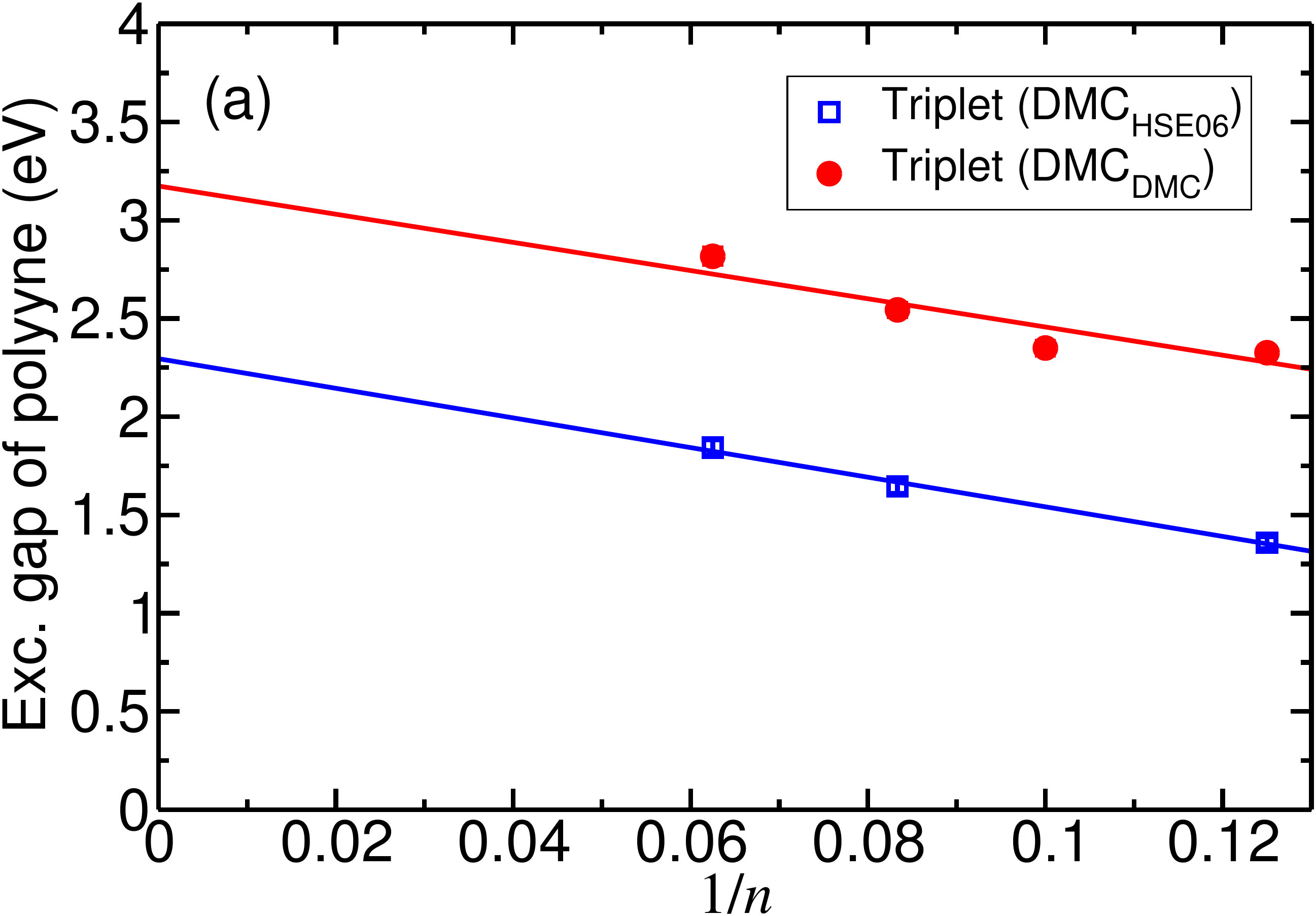}\\ 
\endminipage\hfill
\minipage{0.4\textwidth}
\includegraphics[width=\textwidth]{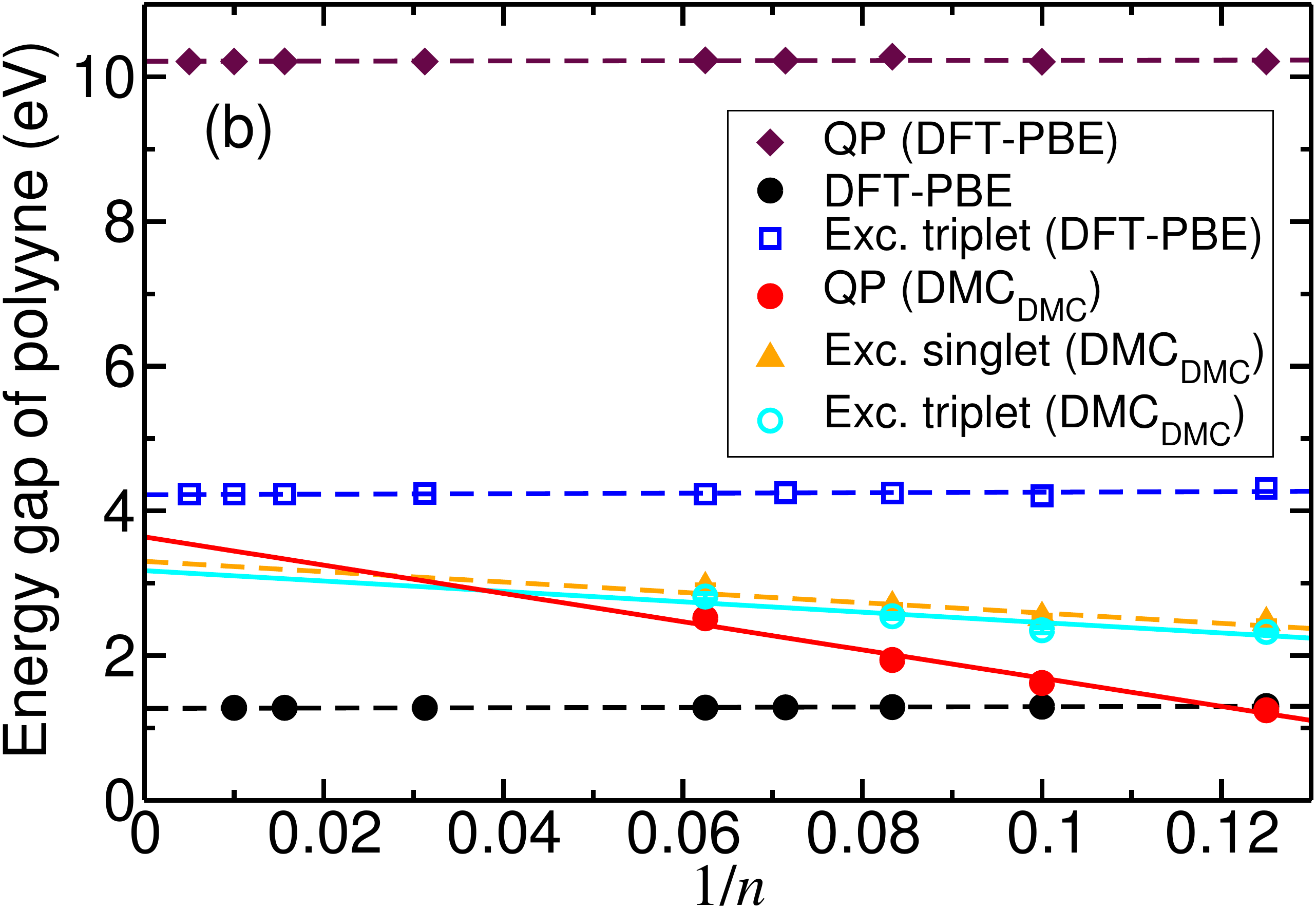}
\endminipage
\caption{(a) DMC excitonic gaps of polyyne against the reciprocal of the
  number $n$ of primitive cells in the supercell as calculated using the
  DFT-HSE06 and the DMC ground-state geometries (DMC$_{\rm HSE06}$ and
  DMC$_{\rm DMC}$, respectively). (b) Quasiparticle (QP) and excitonic energy
  gaps of polyyne against the reciprocal of the number $n$ of primitive cells
  in the supercell as obtained using different methods. The results simply
  labelled ``DFT-PBE'' show the band gap obtained in a ground-state
  band-structure calculation.  The results labelled DMC$_{\rm DMC}$ used the
  DMC ground-state geometry, whereas the results labelled DMC$_{\rm DMC(all)}$
  used the DMC geometries for the ground state, cationic state, and anionic
  state of a finite cell of polyyne when calculating the quasiparticle gap.
  The DFT calculations used the DMC geometries in the same way as the DMC
  calculations. At finite size the quasiparticle gap is smaller than the
  excitonic gap due to the introduction of a neutralising background when a
  charged particle is added to or removed from a periodic cell, as explained
  in Sec.\ \ref{sec:fs_effects}.
  \label{fig:gap_polyyne}}
\end{figure}

Figure \ref{fig:gap_polyyne}(a) shows the finite-size behaviour of the DMC
static-nucleus triplet excitonic gaps of polyyne obtained using the DFT-HSE06
and DMC ground-state geometries. In the infinite-system limit, the DMC triplet
gaps with the DFT-HSE06 and DMC geometries are $2.29 (7)$ and $3.17 (7)$ eV,
respectively. Figure \ref{fig:gap_polyyne}(b) shows the static-nucleus triplet
and singlet excitonic gaps and the quasiparticle gap of polyyne calculated
using the Ewald interaction and the DMC-optimised geometry in different
supercells, together with DFT-PBE gaps. The singlet excitonic gap of polyyne
is slightly larger than the triplet gap. The DFT-PBE quasiparticle and
excitonic gaps are calculated using the DMC-optimised geometry and
Eqs.\ (\ref{eq:gap_quasi}) and (\ref{eq:gap_excitonic}) at different $k$-point
samplings (which may be unfolded to correspond to supercells of $n$ primitive
cells). The triplet excitonic gap calculated by DFT is relatively close to the
DMC triplet excitonic gap, while the DFT quasiparticle gap is far too large.
The DFT gap predicted by the ground-state band-structure calculation is (as
expected) significantly underestimated. The fluctuations in the DFT gaps as a
function of supercell size (i.e., $k$-point grid) are small, suggesting that
single-particle errors in the DMC gaps are negligible.  However, it is clear
that there is a systematically varying finite-size error in the DMC gap. We
have reduced the systematic finite-size errors in our DMC gaps by calculating
both excitonic and quasiparticle gaps for supercells composed of 8, 10, 12,
and 16 primitive cells and then extrapolating to infinite cell size using
Eq.\ (\ref{eq:gap_extrapolated}). The finite-size errors in the quasiparticle
gaps are larger than the finite-size errors in the excitonic gaps, as
discussed in Sec.\ \ref{sec:fs_effects}. The DMC singlet and triplet excitonic
gaps of polyyne calculated using the DMC-relaxed geometry are $3.30(7)$ and
$3.17(7)$ eV, respectively, while the DMC quasiparticle gap is $3.6(1)$ eV\@.

\begin{figure}[!htb]
\begin{center}
\includegraphics[height=5cm]{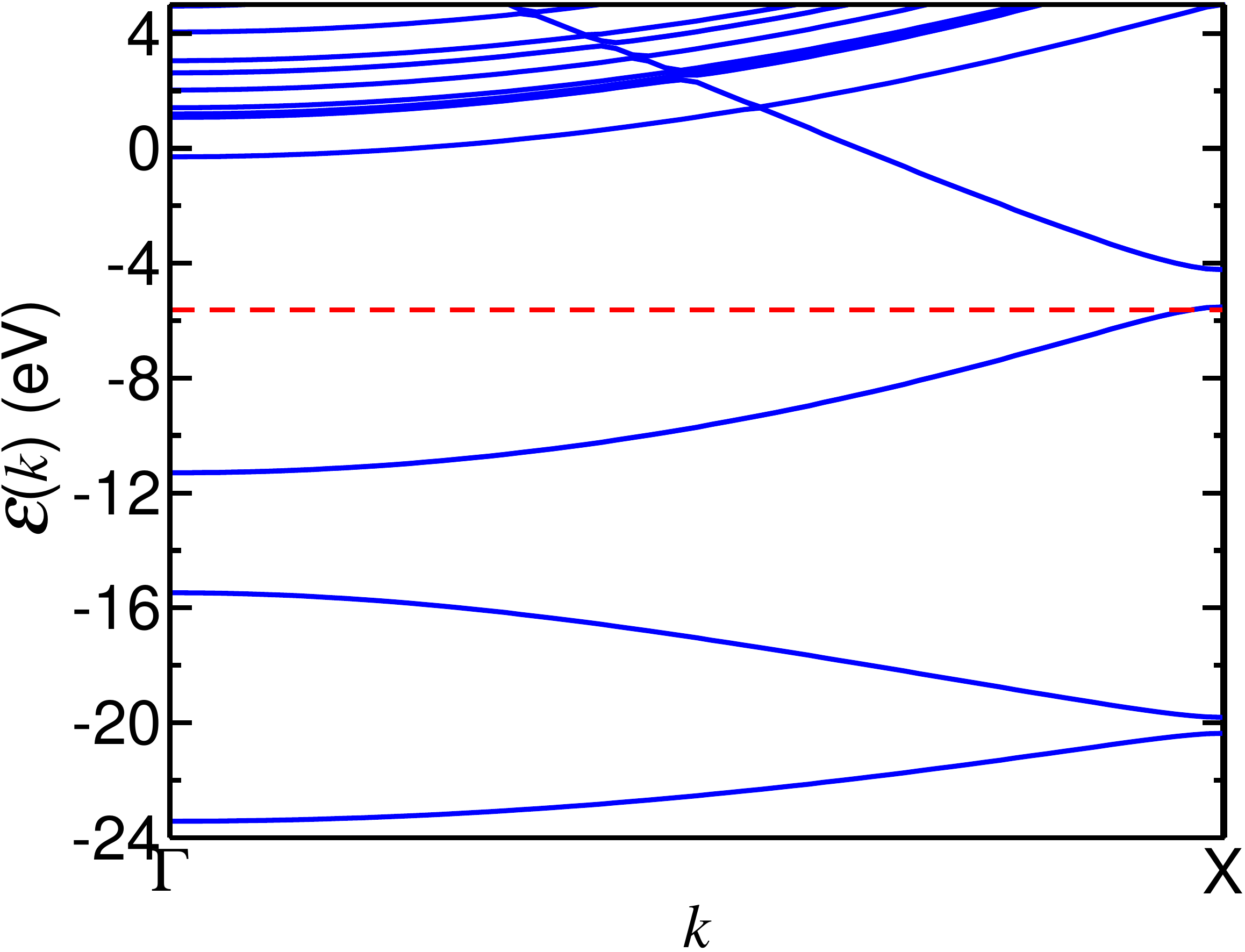}
\caption{DFT-HSE06 band structure of polyyne. The dashed line shows the Fermi
  energy.
\label{fig:bs_hse}}
\end{center}
\end{figure}

To estimate the unscreened exciton binding energy within the Wannier--Mott
model, we have calculated the DFT-HSE06 band structure of polyyne (shown in
Fig.\ \ref{fig:bs_hse}).  In Hartree atomic units the band effective masses
$m_{\rm e}^\ast$ and $m_{\rm h}^\ast$ of the electrons and holes at the $X$
point of the Brillouin zone are given by
\begin{equation} m_{\rm e(h)}^*=\left|\frac{1}{\left(d^2{\cal E}_{\rm
      C(V)}/dk^2\right)_X}\right|, \end{equation} where ${\cal E}_{\rm C}(k)$
and ${\cal E}_{\rm V}(k)$ are the conduction and valance bands,
respectively. Numerically differentiating the DFT-HSE06 bands, we find that
$m_{\rm e}^*=0.046$ a.u.\ and $m_{\rm h}^\ast=0.050$ a.u. In Hartree atomic
units the exciton Bohr radius is $a_0^\ast=1/\mu^\ast$, where $\mu^\ast=m_{\rm
  e}^* m_{\rm h}^*/(m_{\rm e}^*+m_{\rm h}^*)$ is the reduced mass of the
electron--hole pair and we have assumed that the electron and hole interact
via the unscreened Coulomb interaction. In this case, the exciton Bohr radius
is $a_0^\ast=22$ {\AA}, which is slightly smaller than the exciton Bohr radii
of about $30$ \AA\@ estimated for various other 1D conjugated polymers
\cite{Heeger_2001}, and is similar to or smaller than the lengths of the
simulation cells used in our calculations (21--41 {\AA}). Within the
Wannier--Mott model, the unscreened exciton binding energy of polyyne is $1 \,
R_\infty^\ast=\mu^\ast/2 = 0.3$ eV\@. In fact we find the DMC static-nucleus
exciton binding energy to be $0.3(1)$ eV, which is consistent with the small
measured exciton binding energies of a range of $\pi$-conjugated polymers
\cite{knupfer_2003,Moses_2001}.

\begin{figure}[!ht]
\centering
\includegraphics[height=5cm]{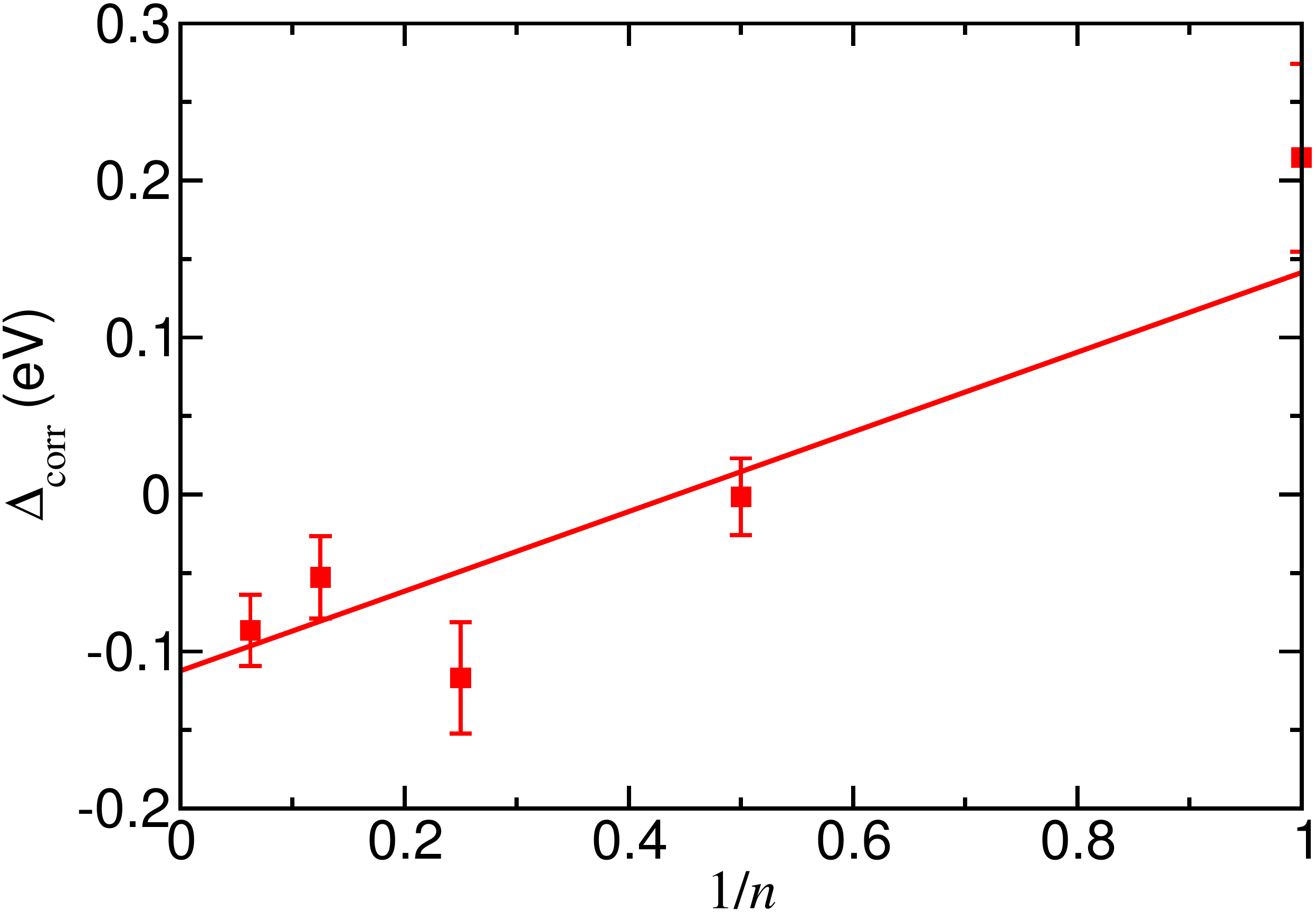}
\caption{DFT-HSE06 zero-point correction $\Delta_{\rm corr}$ to the excitonic
  gap of polyyne against the reciprocal of the number $n$ of primitive cells
  in the supercell.
\label{fig:polyyne_zp}}
\end{figure}

In Fig.\ \ref{fig:polyyne_zp} we report the DFT-HSE06 zero-point correction to
the excitonic gap of polyyne, calculated at different supercell sizes. The
zero-point correction linearly extrapolated to the thermodynamic limit is
$-0.11(2)$ eV\@. As observed for oligoynes, the vibrational correction to the
gap is not as large as in benzene.

In Table \ref{table:compare_oligoynes}, we compare the quasiparticle and
excitonic gaps of polyyne obtained by different methods. The spread of
theoretical results in the literature is remarkable. The static-nucleus DMC
gaps were calculated using the DMC ground-state geometry. The DMC
static-nucleus singlet excitonic gap is $3.30(7)$ eV, which is slightly
reduced to $3.19(7)$ eV\@ by phonon renormalisation.  By extrapolating
experimental absorption gaps of oligoynes to infinite chain length, various
estimates of the gap of polyyne have been made, ranging from 1.24--2.56
eV\@. These are lower than our DMC excitonic gap by $0.63$--$1.95$ eV\@.  We
note that experimental gaps are strongly affected by finite chain length,
solvent, and terminal groups, and that the more recent experimental results on
longer oligoynes (e.g., Ref.\ \citenum{Chalifoux_2010}) are closer to our
results.

\begin{table}[!ht]
\small
\caption{Singlet excitonic gaps $\Delta_{\rm exc}$ and quasiparticle gaps
  $\Delta_{\rm qp}$ of polyyne obtained by different methods. Most of the gaps
  were obtained by extrapolation from a series of oligoyne molecules; the
  number $n$ of pairs of carbon atoms in the largest oligoyne considered in
  each work is shown where known. The DFT-LDA and DFT-BLYP calculations for
  polyyne using periodic boundary conditions (PBC) were performed using 133
  $k$ points \cite{Yang_2006}. Where a citation is not given in the table, the
  data were obtained in the present work.
\label{table:compare_oligoynes}}
\centering
\begin{tabular*}{0.35\textwidth}{@{\extracolsep{\fill}}lcr@{}lr@{}lr@{}l}
\hline 
Method & $n$ & \multicolumn{2}{c}{$\Delta_{\rm exc}$ (eV)}
&\multicolumn{2}{c}{$\Delta_{\rm qp}$ (eV)} \\ \hline
DFT-LDA \cite{Yang_2006}      & PBC & &   & ~~$0.$&$246$\\
DFT-LDAx \cite{Weimer_2005}    & 20 & &   & $0.$&$70$ \\
DFT-PW91 \cite{Yao_2008}       & PBC & &   & $1.$&$17$ \\
DFT-PBE                       & PBC & &  & $1.$&$277$ \\
DFT-PBE1PBE \cite{Yang_2006}   & 36 & &   & $1.$&$801$\\
DFT-B88 \cite{Weimer_2005}     & 20 & &   & $0.$&$72$ \\
DFT-HF \cite{Weimer_2005}      & 20 & &   & $6.$&$31$ \\
DFT-HF \cite{Yang_2006}        & 36 & &   & $8.$&$500$ \\
DFT-LHF \cite{Weimer_2005}     & 20 & &   & $0.$&$92$ \\
DFT-BLYP \cite{Weimer_2005}    & 20 & &   & $0.$&$72$ \\
DFT-BLYP \cite{Yang_2006}     & PBC & &   & $0.$&$320$\\
DFT-B3LYP \cite{Zhuravlev_2004} & 13 & &   & $1.$&$49$\\
DFT-B3LYP \cite{Weimer_2005}   & 20 & &   & $1.$&$50$ \\
DFT-B3LYP \cite{Yang_2006}     & 36 & &   & $1.$&$487$\\
DFT-B3LYP \cite{Peach_2007}    & 12 & &   & $1.$&$59$\\
DFT-KMLYP \cite{Yang_2006}     & 36 & &   & $4.$&$438$\\
DFT-BHHLYP \cite{Yang_2006}    & 36 & &   & $3.$&$946$\\
DFT-BHHLYP \cite{Peach_2007}   & 12 & &   & $4.$&$04$\\
DFT-O3LYP \cite{Yang_2006}     & 36 & &   & $0.$&$895$\\
DFT-CAM-B3LYP \cite{Peach_2007} & 12 & & & $4.$&$33$\\
DFT-HSE06                     & PBC & & & $1.$&$301$\\
$GW$ \cite{Cretu_2013}     & PBC & &   & $0.$&$407$\\
$GW$ \cite{Al-backri_2014} & PBC & &   & $2.$&$15$ \\
MP2 \cite{Yang_2006}       & 20 & &   & $5.$&$541$\\
DMC$_{\rm DMC}$         & PBC &$3.$ &$19(7)$& $3.$&$6(1)$ \\
Experiment \cite{Dembinski_2000} & 10 & $2.$&$20$   &     &  \\
Experiment \cite{Gibtner_2002}   & 10 & $2.$&$20$&  &     \\
Experiment \cite{Mohr_2003}      & 12 & $2.$&$18$--$2.36$ &  & \\
Experiment \cite{Zhuravlev_2004} & 10 & $2.$&$33$&  &     \\
Experiment \cite{Eisler_2005}    & 10 & $2.$&$18$&  &     \\
Experiment \cite{Zheng_2006}     & 12 & $2.$&$16$   &     & \\
Experiment \cite{Samoc_2008}     & 12 & $1.$&$24$--$1.88$ &  & \\
Experiment \cite{Chalifoux_2010} & 22 & $2.$&$56$   &     & \\
\hline 
\end{tabular*}
\end{table}

\section{Conclusions}
In summary we have used DMC to calculate the BLA together with the
quasiparticle and excitonic gaps of hydrogen-capped oligoynes and extended
polyyne.  We have found that simpler levels of theory, such as DFT, do not
predict either the BLA or the gap with quantitative accuracy. Our DMC
calculations show the Peierls-induced BLA of polyyne to be $0.136(2)$ {\AA},
which is significantly higher than DFT predictions.  The DMC quasiparticle gap
of extended polyyne obtained using the DMC-optimised BLA is $3.6(1)$ eV\@.
The static-nucleus DMC singlet excitonic gap of polyyne is $3.30(7)$
eV\@. Vibrational contributions reduce the excitonic gap of polyyne by about
$0.1$ eV\@.  The DMC-calculated zone-centre LO phonon frequency of polyyne is
$2084(5)$ cm$^{-1}$, which is significantly higher than those obtained by DFT,
but is consistent with experimental Raman measurements.  Our work represents
the first direct evaluation of the structural and electronic properties of
extended 1D carbon chains using a high-accuracy method.


\section*{Acknowledgements}
We acknowledge financial support from the UK Engineering and Physical Sciences
Research Council (EPSRC)\@. B.M.\ thanks Robinson College, Cambridge, and the
Cambridge Philosophical Society for a Henslow Research Fellowship. This work
made use of the facilities of Lancaster University's High-End Computing
facility and N8 HPC provided and funded by the N8 consortium and EPSRC
(Grant No.\ EP/K000225/1).


\bibliography{chain_bib}
\end{document}